\documentclass[pdf,final,3p,times,authoryear]{elsarticle}
\usepackage{graphicx}
\usepackage{booktabs}
\usepackage{url}
\usepackage{multicol}
\usepackage{acronym}

\usepackage{xspace}
\usepackage{amsmath}
\usepackage{longtable}
\usepackage{subfigure}

\newcommand{\NA}{\ensuremath{\mathit{NA}}} 
\newcommand{\NAF}{\ensuremath{\mathit{NA}.F}} 
\newcommand{\NAA}{\ensuremath{\mathit{NA}.A}} 
\newcommand{\PQ}{\ensuremath{\mathit{PQ}}} 
\newcommand{\PQF}{\ensuremath{\mathit{PQ}.F}} 
\newcommand{\PQA}{\ensuremath{\mathit{PQ}.A}} 
\newcommand{\PQO}{\ensuremath{\mathit{PQO}}} 
\newcommand{\PQU}{\ensuremath{\mathit{PQU}}} 
\newcommand{\PQOF}{\ensuremath{\mathit{PQO}.F}} 
\newcommand{\PQOA}{\ensuremath{\mathit{PQO}.A}} 
\newcommand{\PQUF}{\ensuremath{\mathit{PQU}.F}} 
\newcommand{\PQUA}{\ensuremath{\mathit{PQU}.A}} 
\newcommand{\PVR}{\ensuremath{\mathit{PVR}}} 
\newcommand{\PVRF}{\ensuremath{\mathit{PVR}.F}} 
\newcommand{\PVRA}{\ensuremath{\mathit{PVR}.A}} 

\newenvironment{fivenum}{
  \begin{tabular}{lllll}
    \toprule
    {\em Min.} & {\em 1st Qu.} & {\em Median} & {\em 3rd Qu.} & {\em Max.} \\   
}{%
  \bottomrule
  \end{tabular}
}

\acrodef{ANVUR}{National Agency for the Assessment of Universities and Research}
\acrodef{ASN}{National Scientific Qualification}
\acrodef{CDF}{Cumulative Distribution Function}
\acrodef{CI}{confidence interval}
\acrodef{MIUR}{Ministry of University and Research}
\acrodef{PVR}{Pareto violation ratio}
\acrodef{SD}{Scientific Discipline}

\acrodef{MCS}{Mathematics and Computer Sciences}
\acrodef{PHY}{Physics}
\acrodef{CHE}{Chemistry}
\acrodef{EAS}{Earth Sciences}
\acrodef{BIO}{Biology}
\acrodef{MED}{Medical Sciences}
\acrodef{AVM}{Agricultural Sciences and Veterinary Medicine}
\acrodef{CEA}{Civil Engineering and Architecture}
\acrodef{IIE}{Industrial and Information Engineering}
\acrodef{APL}{Antiquities, Philology, Literary Studies, Art History}
\acrodef{HPP}{History, Philosophy, Pedagogy and Psychology}
\acrodef{LAW}{Law}
\acrodef{ECS}{Economics and Statistics}
\acrodef{PSS}{Political and Social Sciences}

\begin{document}

\begin{frontmatter}

\title{Quantitative Analysis of the Italian National Scientific Qualification\tnoteref{label1}}
\tnotetext[label1]{The final version of this paper is published in the Journal of Informetrics. Please cite as: Moreno Marzolla, \emph{Quantitative Analysis of the Italian National Scientific Qualification}, Journal of Informetrics 9(2), April 2015, pp. 285--316, ISSN 1751-1577, DOI \url{http://dx.doi.org/10.1016/j.joi.2015.02.006}}
\author{Moreno Marzolla\corref{cor1}}
\ead{moreno.marzolla@unibo.it}
\cortext[cor1]{Corresponding Author, Department of Computer Science and Engineering, University of Bologna, mura Anteo Zamboni 7, I-40126 Bologna, Italy. Phone +39 051 20 94847, Fax +39 051 20 94510}
\address{Department of Computer Science and Engineering, University of Bologna, Italy}

\begin{abstract}
  The Italian National Scientific Qualification (ASN) was introduced
  in 2010 as part of a major reform of the national university
  system. Under the new regulation, the scientific qualification for a
  specific role (associate or full professor) and field of study is
  required to apply for a permanent professor position. The~ASN is
  peculiar since it makes use of bibliometric indicators with
  associated thresholds as one of the parameters used to assess
  applicants. The first round of the~ASN
  received~59,149 applications,
  and the results have been made publicly available for a short period
  of time, including the values of the quantitative indicators for
  each applicant. The availability of this wealth of information
  provides an opportunity to draw a fairly detailed picture of a
  nation-wide evaluation exercise, and to study the impact of the
  bibliometric indicators on the qualification results. In this paper
  we provide a first account of the Italian National Scientific
  Qualification from a quantitative point of view. We show that
  significant differences exist among scientific disciplines, in
  particular with respect to the fraction of qualified applicants,
  that can not be easily explained. Furthermore, we describe some
  issues related to the definition and use of the bibliometric
  indicators and the corresponding thresholds. Our analysis aims at
  drawing attention to potential problems that should be addressed by
  decision-makers in future rounds of the~ASN.
\end{abstract}

\begin{keyword}
  National Scientific Qualification; ASN; Research Evaluation; Bibliometrics; Italy
\end{keyword}

\end{frontmatter}

\section{Introduction}

The Italian Law 240/2010~\citep{L240} introduced substantial changes
in the national university system. The law is quite broad in scope: it
requires universities to undergo a major internal reorganization,
delegates the government to define new rules for improving the quality
and efficiency of higher education system, and modifies the
recruitment process of university professors. Under the new
regulation, to apply for a permanent professor positions it is first
necessary to acquire the~\ac{ASN}\footnote{The acronym ASN stands for
  \emph{Abilitazione Scientifica Nazionale}. All acronyms used in this
  paper (e.g., ASN, MIUR, ANVUR) are based on the original (Italian)
  denomination, since they have a well established meaning for the
  Italian scientific community, while the expanded forms are in
  English for the benefit of international readers.}. The~\ac{ASN} is
meant to attest that an individual has reached the scientific maturity
required for applying for a specific role (associate or full
professor) in a given scientific discipline~\cite[Art. 16]{L240};
however, the qualification does not guarantee that a professorship
position will eventually be granted.

The Italian~\ac{ASN} is similar to other \emph{habilitation}
procedures already in place in other countries (e.g.,~France and
Germany) in that it is a prerequisite for becoming a university
professor. What makes the~\ac{ASN} peculiar is its reliance on
bibliometry as one of the parameters used to evaluate applicants.
Specifically, the Ministry of University and Research defined three
quantitative indicators whose values were computed for each
candidate. To grant qualification, examination committees should take
into account how many indicators exceed previously computed
thresholds. Such thresholds were defined as the medians of the values
of those indicators for tenured professors.

The first round of the~\ac{ASN} started on November 2012 and completed
in August 2014 with the publication of the last batch of
results. Given the stagnating status of the Italian university system,
the~\ac{ASN} represents the only opportunity for postdocs and
temporary researchers to move towards a permanent position, and for
tenured researchers and associate professors to move up the academic
ladder. Therefore, it is not surprising that the~\ac{ASN}
received~59,149 applications
spanning 184 scientific disciplines. The curricula of all applicants,
the values of their bibliometric indicators and the final reports of
examination committees have been made publicly available for a short
period of time. This provided an opportunity to analyze a nation-wide
research assessment procedure involving a large number of applicants
from all scientific areas.

The present work describes the results of the Italian~\ac{ASN} from a
quantitative point of view. Specifically, we compute several
statistics that provide a picture of the outcome of the qualification
procedure. These statistics include: the fraction of successful
qualifications, whether the values of bibliometric indicators are
correlated with the outcome, whether those indicators have been used
consistently across applicants, and whether the values of the
quantitative indicators are correlated with the qualification result.

This paper is descriptive and aims at showing what happened in order
to provide insights and draw attention to potential problems that
require further investigation. Although we try to suggest possible
explanations whenever appropriate, it is understood that only manual
examination of the applicants' curricula and final reports may reveal
whether the problems highlighted here are real issues. The~\ac{ASN}
has been criticized by part of the Italian scientific community as a
form of ``career assessment by numbers'' (using a term borrowed
from~\cite{KellyJe06}), due to its reliance on (ad-hoc) bibliometric
indicators for individual evaluation. In this paper we try to avoid
using numbers to explain what must be left to human judgment.

This paper is organized as follows. In Section~\ref{sec:overview} we
provide the necessary background on the structure of the Italian
university system and the~\acl{ASN} rules. In
Section~\ref{sec:data-collection} we give a high level overview of the
outcome of the~\ac{ASN} in terms of the number of applications and
percentages of successful qualifications for each scientific area and
discipline. Then, we turn our attention to the numerical indicators
used to evaluate the applicants: the thresholds (medians) will be
studied in Section~\ref{sec:medians}, while the impact of the
bibliometric indicators on the outcome of the qualification procedure
will be analyzed in Section~\ref{sec:bib-indicators}. Final discussion
and concluding remarks are presented in Section~\ref{sec:conclusions}.

\section{The Italian National Scientific Qualification}\label{sec:overview}

\subsection{Overview}

Before 2010, there were three types of tenured research and teaching
positions at Italian universities: assistant professor, associate
professor and full professor.  Law 240/2010 replaced the role of
tenured assistant professor with two types of fixed term positions,
called \emph{Type A} and \emph{Type B} researcher. Type A positions
last for three years and can be extended once for two more years,
while type B positions last for three years with no provision for
further extension. Type A positions are supposed to be a path towards
becoming tenured associate professor, since universities hiring type A
researchers must allocate funding for promotion in advance.

\begin{table}[t]
\centering\begin{tabular*}{\textwidth}{@{\extracolsep{\fill}}llll}
\toprule
{\em ID} & {\em Code} & {\em Area Name} & {\em N. of Sc. Disciplines} \\
\midrule
01 & MCS & \acl{MCS} &   7 \\
02 & PHY & \acl{PHY} &   6 \\
03 & CHE & \acl{CHE} &   8 \\
04 & EAS & \acl{EAS} &   4 \\
05 & BIO & \acl{BIO} &  13 \\
06 & MED & \acl{MED} &  26 \\
07 & AVM & \acl{AVM} &  14 \\
08 & CEA & \acl{CEA} &  12 \\
09 & IIE & \acl{IIE} &  20 \\
10 & APL & \acl{APL} &  19 \\
11 & HPP & \acl{HPP} &  17 \\
12 & LAW & \acl{LAW} &  16 \\ 
13 & ECS & \acl{ECS} &  15 \\
14 & PSS & \acl{PSS} &   7 \\
\midrule 
   & Total &         & 184 \\
\bottomrule
\end{tabular*}
\caption{The 14 scientific areas. For each area we show the numeric
  ID, a three-letter code, the area name and number of scientific
  disciplines it contains.}\label{tab:area}
\end{table}

Each professor and researcher is bound to a specific field of study,
called~\acf{SD}. When the~\ac{ASN} started, scientific disciplines
were organized in 14 \emph{scientific areas}, each one comprising
several macro-sectors that were further divided into~\acp{SD}.  The~14
scientific ares are listed in Table~\ref{tab:area}; for each area we
show its numeric ID, a three-letter acronym, the name and number
of~\acp{SD} it contains. Overall, 184~\acp{SD} were
defined~\cite[Annex A]{dm-set-concorsuali}; for completeness, they are
listed in~\ref{app:list-sd}. The aim and scope of each discipline is
given in~\cite[Annex B]{dm-set-concorsuali}. Each~\ac{SD} is
identified by a four-character code of the form \emph{AA/MC} where
\emph{AA} is the ID of the area the discipline belongs to (01--14),
\emph{M} is a single letter denoting the macro-sector, and \emph{C} is
a digit identifying the discipline within the macro-sector. For
example, 01/B1 denotes Computer Science, 09/H1 denotes Computer
Engineering and 11/A1 denotes Medieval History.

Each university can create new positions for a given~\ac{SD} and role
(associate or full professor), provided that certain administrative
and financial requirements are met. Once the position is advertised,
only those that have acquired the~\ac{ASN} for that specific~\ac{SD}
and role can apply. In this paper we use the term~\ac{ASN} to denote
both the qualification procedure used to grant qualification, and the
qualification itself. It is important to observe that a qualification
does not, by itself, guarantee any position, but merely allows the
owner to apply for a professorship; each university handles the hiring
process according to locally defined rules.

The~\ac{ASN} is supposed to be held once a year. For each~\ac{SD},
the~\ac{MIUR} appoints a five-member examination committee which is in
charge of evaluating all applications, both at the full and associate
professor levels, Committee members are randomly selected from a list
of eligible professors. To be eligible, one has to satisfy
quantitative requirements similar to those used for assessing
applicants (see below). Each committee is made of four full professors
from Italian universities, and one professor from a foreign university
or research institution. Therefore, 920 examiners were
appointed; it is worth noticing that each of the 184 foreign
professors was paid Eur 16,000 for two years~\citep{dm159}, for a total
cost of Eur 2,944,000.

Once acquired, a qualification lasts for six years; those who fail to
get qualification can not apply again for two years.  Although the
scientific qualification is bound to a specific~\ac{SD} (e.g.,
computer science) and role (e.g., associate professor), it was
possible to apply for multiple qualifications for different~\acp{SD}
and/or role. Indeed, there are many applicants who got multiple
qualifications; one particularly successful researcher acquired 8
qualifications for both the full and associate professor levels, in 8
related scientific disciplines. There has also been the curious
situations of one applicant who successfully qualified as full
professor in discipline 06/H1--\emph{Obstetrics and gynecology}, but
was denied qualification as associate professor in the same discipline
(by the same examination committee)!  This was probably due to an
error in the submitted application form for associate professor
qualification, since a significant number of publications were
missing.

\subsection{Quantitative Indicators}\label{sec:indicators}

Law 240/2010 made the provision that the~\ac{ASN} has to be granted
based on the analytic evaluation of scientific publications using
criteria and parameters defined in a separate decree. Those parameters
were eventually described in the Ministerial Decree
76/2012~\citep{dm76} that introduced two flavors of quantitative
indicators, called \emph{bibliometric} and \emph{non-bibliometric}
indicators, respectively.

Bibliometric indicators~\cite[Annex A]{dm76} apply to scientific
disciplines for which ``sufficiently complete'' citation databases
exist. The three bibliometric indicators are the following:

\begin{enumerate}
\item[$B_1$] Normalized number of journal papers;
\item[$B_2$] Total normalized number of citations received;
\item[$B_3$] Normalized $h$-index.
\end{enumerate}

These are used for all disciplines belonging to the nine scientific
areas~\ac{MCS}, \ac{PHY}, \ac{CHE}, \ac{EAS}, \ac{BIO}, \ac{MED},
\ac{AVM}, \ac{CEA} and \ac{IIE}, with the exception of
08/C1--\emph{Design and technological planning of architecture},
08/D1--\emph{Architectural design}, 08/E1--\emph{Drawing},
08/E2--\emph{Architectural restoration and history} and
08/F1--\emph{Urban and landscape planning and design}, but including
the whole macro sector 11/E--\emph{Psychology}. These disciplines are
denoted as \emph{bibliometric disciplines}.

Non-bibliometric indicators~\cite[Annex B]{dm76} apply to scientific
disciplines (mostly, humanities and social sciences) for which
citation-based indices can not be reliably computed due to the lack of
suitable bibliometric databases. Non-bibliometric indicators have been
defined for~\ac{APL}, \ac{HPP}, \ac{LAW}, \ac{ECS} and \ac{PSS}, with
the exceptions above; these disciplines are referred to as
\emph{non-bibliometric disciplines} in the official~\ac{MIUR}
documents. The three non-bibliometric indicators are:

\begin{enumerate}
\item[$N_1$] Normalized number of published books;
\item[$N_2$] Normalized number of journal papers and book chapters;
\item[$N_3$] Normalized number of papers published on ``top'' journals.
\end{enumerate}

It is important to remark that the terms ``bibliometric'' and
``non-bibliometric'' are used in the official~\ac{MIUR}
documents. Unfortunately, such terminology is not consistent with that
used by the scientometric community, since the so-called
non-bibliometric indicators $N_1$, $N_2$ and $N_3$ are indeed
\emph{bibliometric}, being based on paper counts. Given that the terms
``bibliometric'' and ``non-bibliometric'' became standard within the
Italian research community, we will follow the~\ac{MIUR} ``newspeak''
according to the definitions above.  However, to limit further
confusion we will use the generic term \emph{quantitative indicators}
($\mathit{ind}_i$) to refer to both the bibliometric indicators $B_i$
and non-bibliometric ones $N_i$.

Normalization of quantitative indicators is used to mitigate the bias
against young applicants, since citations and paper counting metrics
increase over time. Normalization is based on the concept of
\emph{scientific age}~\cite{anvur-normalizzazione}: an applicant that
published her first paper in year $t_0$ has scientific age
$\mathit{SA}$ equal to:

\begin{equation}
  \mathit{SA} = \max\left\{10, (2012 - t_0 + 1)\right\}\label{eq:sa}
\end{equation}

Indicators $B_1$, $N_1$, $N_2$ and $N_3$ are normalized by multiplying
the raw value by $10 / \mathit{SA}$. Indicator $B_2$ (number of
citations received) is normalized by dividing the raw value by the
academic age. Finally, indicator $B_3$ (normalized $h$-index) requires
a more complex procedure. Given a paper $p$, published in year $t_p$,
that at time $t \geq t_p$ has received $C(p, t)$ citations, the
normalized number of citations $S(p, t)$ for $p$ is defined as:

\begin{equation}
  S(p, t) = \frac{4}{t - t_p + 1} C(p, t)
\end{equation}

Then, the normalized h-index $h_c$ is defined as the maximum integer
such that $h_c$ papers of a given applicant received at least $h_c$
\emph{normalized} citations each~\citep{hc-index}.  

The values of the quantitative indicators were computed for each
applicant by~\ac{ANVUR}, a public entity that oversees the evaluation
of universities and public research institutes, using data from Scopus
and ISI Web of Science (WoS). The number of citations of each paper
was computed as the maximum value reported by Scopus and WoS. Only
publications produced during the ten years period 2002--2012 were
considered.

\subsection{Medians}

For each quantitative (bibliometric and non-bibliometric) indicator
\ac{ANVUR} computed a threshold, defined in~\cite{dm76} as the
\emph{median} of the values assumed by the indicator for tenured
associate and full professors. All professors were asked to
voluntarily upload their lists of publications to a central database,
so that the values of the appropriate quantitative indicators could be
computed~\citep{anvur-delibera-mediane}. Separate medians were
computed for each discipline and for each role (full and associate
professor). The thresholds were computed by~\ac{ANVUR} before the
closing date for applicants. Candidates were then informed about the
values of the medians and of their individual quantitative indicators;
those who were no longer willing to pursue their application, i.e.,
because their quantitative indicators were below the medians, could
withdraw from the~\ac{ASN}.

Special provisions were made to cope with~\acp{SD} where the
quantitative indicators exhibited a multi-modal
distribution~\cite[Art. 16]{anvur-delibera-mediane}. This was handled
by defining one or more sub-disciplines with different
medians. However, this procedure could only be applied to~\acp{SD}
with no less than 100 full professors, and required ad-hoc interaction
with the National University
Council. 47
sub-disciplines where identified for full professor,
and~47 for associate
professor qualification (see the table in~\ref{app:medians}).

According to the literal interpretation of~\cite[Art. 6.1, 6.2]{dm76},
only applicants whose quantitative indicators exceed two or one
medians (for bibliometric and non-bibliometric disciplines,
respectively) could get the qualification.  For example, suppose that
the medians for a given bibliometric discipline are $(10, 13.2,
7)$. An applicant whose quantitative indicators are $(11, 15, 6)$
exceeds the first and second median, and according to the
interpretation above may obtain the qualification, provided that other
qualitative aspects of his scientific profile are also evaluated
positively. On the other hand, an applicant with indicators $(13, 12,
7)$ only exceeds the first median (the value of the third indicator is
equal to the corresponding median, and therefore does not exceed it),
so it is not eligible for qualification.

In other words, exceeding one or two medians was initially considered
a necessary but not sufficient condition for qualification. This
interpretation was later relaxed~\citep{circolare-mediane}, also
because~\cite[Art 6.5]{dm76} partially conflicts with~\cite[Art. 6.1,
  6.2]{dm76} by stating that examination committees can waive the
requirement above, provided that their decision is motivated in the
final report. In Section~\ref{sec:bib-indicators} we will investigate
whether committees likely adopted the restrictive or the relaxed
interpretation of the requirement above.

\subsection{Discussion}

Although a detailed discussion is outside the scope of this paper, we
want to point out some methodological issues of the~\ac{ASN}.

\begin{itemize}

\item It is well known that quantitative indicators must be used with
  extreme care when evaluating individual
  researchers~\citep{Sahel11,ieee-bib,LaloeMo09,proper}. The Italian
  lawmaker appears to be aware at least of the most serious pitfalls
  and tried to address them, e.g., using multiple indicators instead
  of one, using normalization to take into account the scientific age
  of applicants, and making the provision that also the qualitative
  profile of applicants had to be taken into account. Some of those
  countermeasures introduce other problems (see below). Moreover, the
  law~\citep{dm76} is ambiguous about the role of quantitative
  indicators, and this generated confusion among applicants and
  examination committees.
  
\item The idea of using the medians of quantitative indicators of
  tenured professors as thresholds is a form of ``grading by curves''
  that leads to bizarre consequences.  By construction, half of the
  tenured professors do not exceed each median; this is true
  regardless of their quality: should they all be Nobel laureates,
  half of them would still be below each threshold. This ensures that
  a fraction of tenured professors will not meet the quantitative
  qualification requirements for \emph{their own} role. The problem is
  not that some professors fail qualification for their current
  position (this may happen with different criteria as well), but that
  this happens by construction.

\item The values of medians provided by~\ac{ANVUR} can not be
  validated, since they were computed using a list of publication that
  has not been made publicly available. Since~\ac{ANVUR} released a
  second set of medians to fix errors in their original computations,
  one may wonder whether the new values are indeed correct.
  
\item The use of Scopus and WoS as the only sources of bibliometric
  information places considerable power on the hands of private
  companies. Some applicants did not have access to these databases
  (e.g., because they did not have an institutional subscription and
  were not able or willing to pay for one), and therefore could not
  verify the correctness of their data. In any case, Scopus and WoS
  were under no obligation to fix errors before the~\ac{ASN} deadline,
  or fix errors at all.
  
\item Although normalization of quantitative indicators tries to
  address a valid concern, i.e., that paper-counting and
  citation-based metrics penalize younger applicants, its
  implementation in the~\ac{ASN} according to Eq.~\eqref{eq:sa}
  creates the so called ``paradox of academic twins''. Consider two
  applicants, Alice and Bob, who have only joint publications over the
  8 years preceding the~\ac{ASN}. Alice has no other publications
  while Bob has one additional conference paper, published ten years
  before~\ac{ASN}, that received no citations. Normalization only
  applies to Alice, since her scientific age is less than ten
  years. Due to normalization, Alice indicators are higher than those
  of Bob, even though she has a strict subset of Bob's
  publications. In general, the values of quantitative indicators can
  be increased by intentional or accidental omission of older
  publications.
    
\end{itemize}

\section{General Overview}\label{sec:data-collection}

The~\ac{ASN} results were made available for three months after the
initial publication at \url{http://abilitazione.miur.it/}. The data
were provided as HTML pages and PDF documents that are appropriate for
manual browsing but not for automatic processing. Therefore, we
developed a crawler that extracted and formatted the relevant
information in CSV (Comma Separated Values) format; all subsequent
analyses have been performed using R~\citep{r-stat}.

The following data have been used in this paper:
\begin{itemize}
  \item Applicant first and last name (string);
  \item Scientific discipline (and optional sub-discipline) applied to
    (string);
  \item Role applied to (integer, 1=full professor, 2=associate professor);
  \item Values of the three quantitative indicators;
  \item Result of the qualification procedure (boolean, \emph{true}=qualified,
    \emph{false}=not qualified);
\end{itemize}

The values of medians for bibliometric and non-bibliometric
disciplines were taken from the official~\ac{ANVUR}
documents~\citep{anvur-mediane-po-bib, anvur-mediane-pa-bib,
  anvur-mediane-po-nonbib, anvur-mediane-pa-nonbib,
  anvur-mediane-area12}.

\begin{table}[t]
\begin{center}
  \begin{small}
    \begin{tabular*}{\textwidth}{@{\extracolsep{\fill}}llllllllll}
      \toprule
          {\em Area} & \multicolumn{3}{l}{\em Full professor} & \multicolumn{3}{l}{\em Associate professor} & \multicolumn{2}{l}{\em Total} \\
          \cmidrule{2-4}\cmidrule{5-7}\cmidrule{8-10}
          & {\em Applications} & {\em Qualified} & $\PQF$ & {\em Applications} & {\em Qualified} & $\PQA$ & {\em Applications} & {\em Qualified} & $\PQ$ \\
          \midrule
 MCS & 911 & 356 & 0.391 & 1581 & 714 & 0.452 & 2492 & 1070 & 0.429 \\ 
  PHY & 1451 & 760 & 0.524 & 2921 & 1676 & 0.574 & 4372 & 2436 & 0.557 \\ 
  CHE & 695 & 387 & 0.557 & 1649 & 934 & 0.566 & 2344 & 1321 & 0.564 \\ 
  EAS & 400 & 148 & 0.370 & 831 & 366 & 0.440 & 1231 & 514 & 0.418 \\ 
  BIO & 1690 & 763 & 0.451 & 4554 & 1874 & 0.412 & 6244 & 2637 & 0.422 \\ 
  MED & 3298 & 1377 & 0.418 & 6689 & 2669 & 0.399 & 9987 & 4046 & 0.405 \\ 
  AVM & 650 & 373 & 0.574 & 1443 & 747 & 0.518 & 2093 & 1120 & 0.535 \\ 
  CEA & 1027 & 371 & 0.361 & 2572 & 906 & 0.352 & 3599 & 1277 & 0.355 \\ 
  IIE & 1573 & 691 & 0.439 & 2962 & 1256 & 0.424 & 4535 & 1947 & 0.429 \\ 
  APL & 1718 & 796 & 0.463 & 4606 & 2082 & 0.452 & 6324 & 2878 & 0.455 \\ 
  HPP & 1491 & 509 & 0.341 & 4418 & 1632 & 0.369 & 5909 & 2141 & 0.362 \\ 
  LAW & 887 & 322 & 0.363 & 2150 & 736 & 0.342 & 3037 & 1058 & 0.348 \\ 
  ECS & 1755 & 787 & 0.448 & 3098 & 1451 & 0.468 & 4853 & 2238 & 0.461 \\ 
  PSS & 515 & 162 & 0.315 & 1614 & 497 & 0.308 & 2129 & 659 & 0.310 \\ \midrule
 Total & 18061 & 7802 & 0.432 & 41088 & 17540 & 0.427 & 59149 & 25342 & 0.428 \\ \bottomrule
\end{tabular*}
\end{small}
\end{center}
\caption{Number of applications for each area and
  role.}\label{tab:dati-aree}
\end{table}

Table~\ref{tab:dati-aree} reports the number of submitted
applications, the number of successful applications and the fraction
of successful applications ($\PQ$) for each area; an application is
successful if it leads to qualification. Variables with the suffixes
$.F$ and $.A$ refer to full and associate roles, respectively.
Therefore, $\PQF$ is the fraction of applicants for the full professor
level that got qualification, and $\PQA$ is the fraction of applicants
for the associate professor level that got qualification.

Overall, 59,149 applications have been
submitted\footnote{The actual number
  is~59,150, but no quantitative indicators
  were shown for one of the applicants, and therefore we dropped that
  entry from this analysis.}, 18,061 for full
and~41,088 for associate professor
qualification. 25,342 applications were
successful (42.8\%). As can be seen, the
fraction of successful qualifications for the full and associate
levels were almost identical.  Note that the number of
\emph{applications} is higher than the number of \emph{applicants},
since many individuals submitted multiple applications. We
counted~39,583
unique $\langle \textit{last name}, \textit{first name} \rangle$
pairs, but the number of individuals may be higher due to the presence
of people with the same names.

Area~\ac{MED} received the
highest number of applications
(9,987), while
area~\ac{EAS} received the lowest
(1,231). The
area with highest percentage of successful qualifications
is~\ac{CHE}
(56.4\%), while the
one with lowest percentage
is~\ac{PSS}
(31\%).
The percentage of successful qualifications for bibliometric
disciplines is~44.6\%, while the
percentage of qualifications for non-bibliometric disciplines
is~40.1\%. The difference is quite
small and does not denote any particular bias.

\begin{figure}[t]
\centering\includegraphics[width=\textwidth]{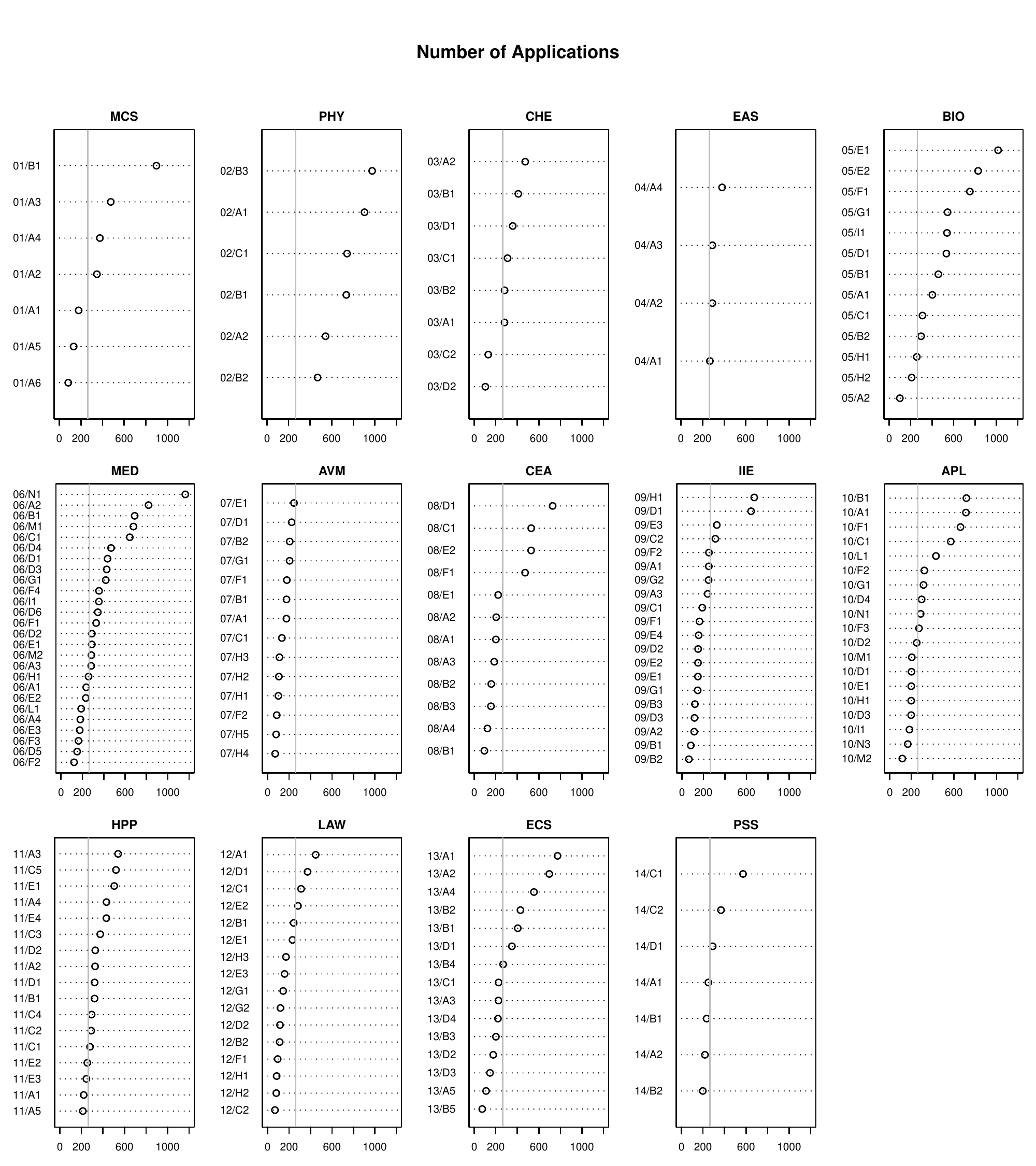}
\caption{Total number of applications for each scientific discipline;
  the vertical lines denote the population median across all
  disciplines (263).}\label{fig:n-applications}
\end{figure}

From Table~\ref{tab:dati-aree} we observe large differences across
areas, both in the number of applications and fraction of successful
qualifications. To better investigate these differences we perform a
more detailed analysis at the level of individual~\acp{SD}.
Figure~\ref{fig:n-applications} shows the number of applications
($\NA$) for each~\acp{SD} (the raw data can be found
in~\ref{app:statistics}). The median number of applications
is~263; 75\% of the~\acp{SD} received less
than~421.2  applications, while the remaining
25\% received up to~1,164.

In this paper we use Tukey's five number
summary~\citep{tukey1977exploratory} (minimum, first quartile, sample
median, third quartile, maximum) to describe the shape of
distributions, since quantile-based statistics are more robust than
sample mean and standard deviation. The five-number summary for the
number of applications $\NA$ is shown in
Table~\ref{fn:num-applications}.

\begin{table}[ht]
  \centering%
  \caption{Number of Applications}\label{fn:num-applications}
  \begin{fivenum}
    64 & 175 & 263 & 421.2 & 1164\\
  \end{fivenum}
\end{table}
    
Considerable variability exists across different~\acp{SD}: the minimum
number of applications was submitted to 09/B2--\emph{Industrial
  mechanical plants} (64 applications, 13 for full and~51 for
associate professor level), while the maximum number of applications
was submitted to 06/N1--\emph{Applied medical technologies} (1164
applications, 365 for full and~799 for associate professor
level). Figure~\ref{fig:n-applications} shows that all disciplines of
area~\ac{PHY} received more applications than the median, while those
of area~\ac{AVM} received less applications than the median.

This information, other than providing a very rough estimate of the
number of researchers in each discipline, is also useful from a
practical point of view. Every examination committee was given the
same amount of time (two months) for processing all applications. If
done properly, this involves the following activities (for each
candidate): (i)~assessing the CV; (ii)~assessing the publications
provided in full text; (iii)~writing the final report. These
activities require a significant amount of time, especially if the
number of applications to process is large; as an example, the
committee for discipline 06/N1--\emph{Applied medical technologies}
was appointed on December 27, 2012
and was supposed to
process~1164
applications by February~25, 2013. Applicants to area~\ac{MED} were
allowed to submit at most 20 publications in full text for full
professor qualification, and 14 for associate professor
qualification. Therefore, each committee member of~\ac{SD} 06/N1 was
supposed to evaluate~1,164 CVs and about~$20 \times 365 + 14 \times
799 = 18,486$ publications in two months. It is
not surprising that~\ac{MIUR} had to grant multiple deadline
extensions to committees that were in a similar
situation\footnote{\emph{Decreto Direttoriale 47, 2012-01-09}
  \url{http://abilitazione.miur.it/public/documenti/commissioni/Proroga_termini_090113.pdf}},
and this produced delays in the publication of final results. Future
rounds of the~\ac{ASN} should take the workload of examination
committees into account, and consider splitting overcrowded~\acp{SD}
across different committees. Extra care should be taken for ensuring
that the evaluations are as much committee-independent as possible.

\begin{figure}[t]
\centering\includegraphics[width=.4\textwidth]{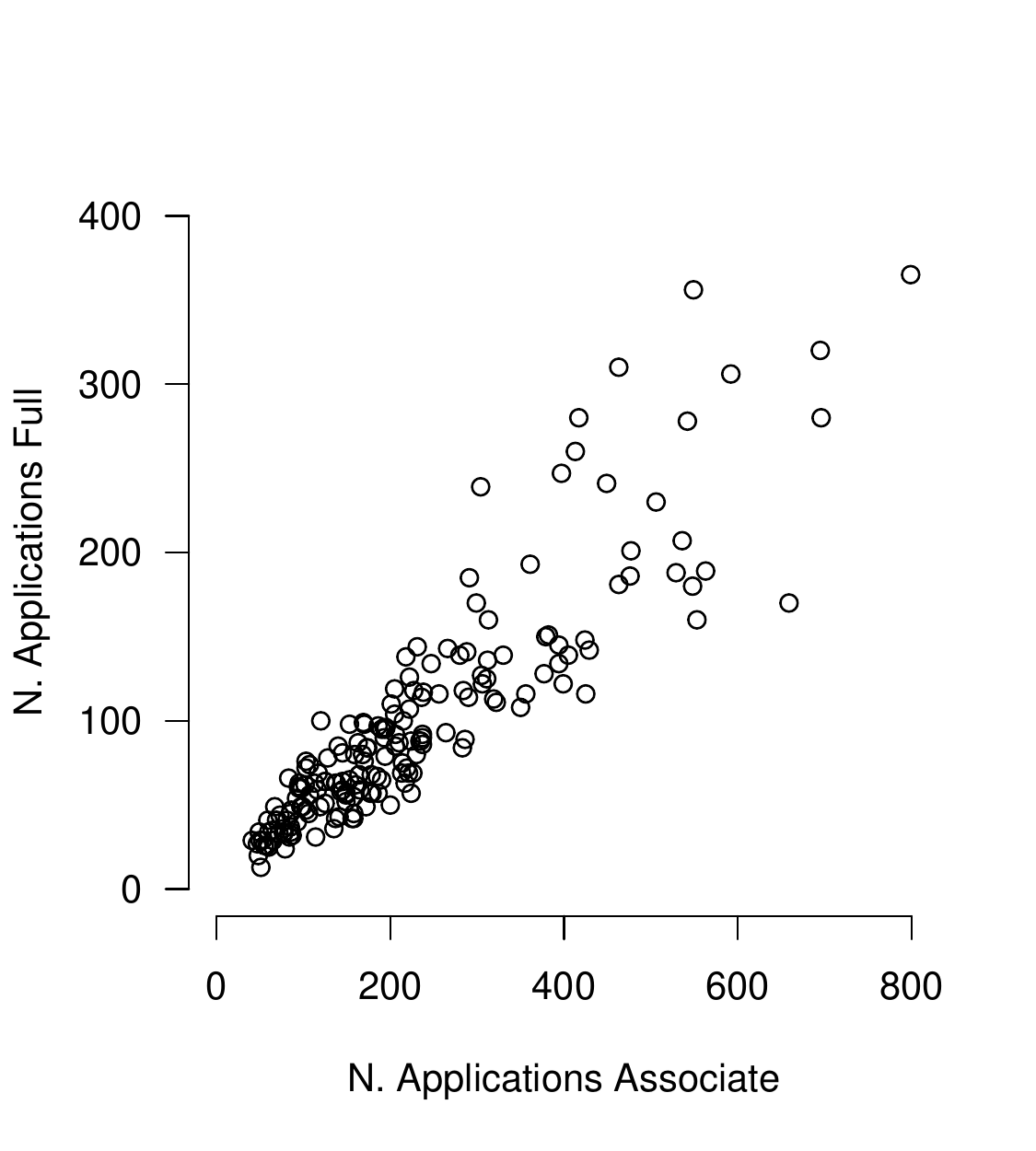}
\caption{Number of applications for the full versus associate professor
  levels. Each data point represents
  a~\ac{SD}.}\label{fig:cor-n-applications}
\end{figure}

The number of applications for the full ($\NAF$) and associate ($\NAA$)
levels are strongly correlated. We use Spearman's rank order
correlation coefficient $\rho$ to measure the strength of the
correlation~\citep{myers2010research}. Spearman's $\rho$ is a
non-parametric measure of association between two samples; values
closer to 1 denote higher (positive) correlation. We prefer Spearman's
$\rho$ to the more commonly used Pearson's product-moment correlation
coefficient $\tau$ since the latter only measures linear correlation,
while Spearman's $\rho$ estimates how well the dependency of two
variables can be described using a generic monotonic function. The
correlation coefficient for the number of applications at the full
versus associate professor level is $\rho(\NAF, \NAA) = 0.91$, denoting significant positive correlation
(see Figure~\ref{fig:cor-n-applications}); the 95\%~\ac{CI} for the
correlation coefficient is $[0.88, 0.94]$.

\begin{figure}[t]
\centering\includegraphics[width=\textwidth]{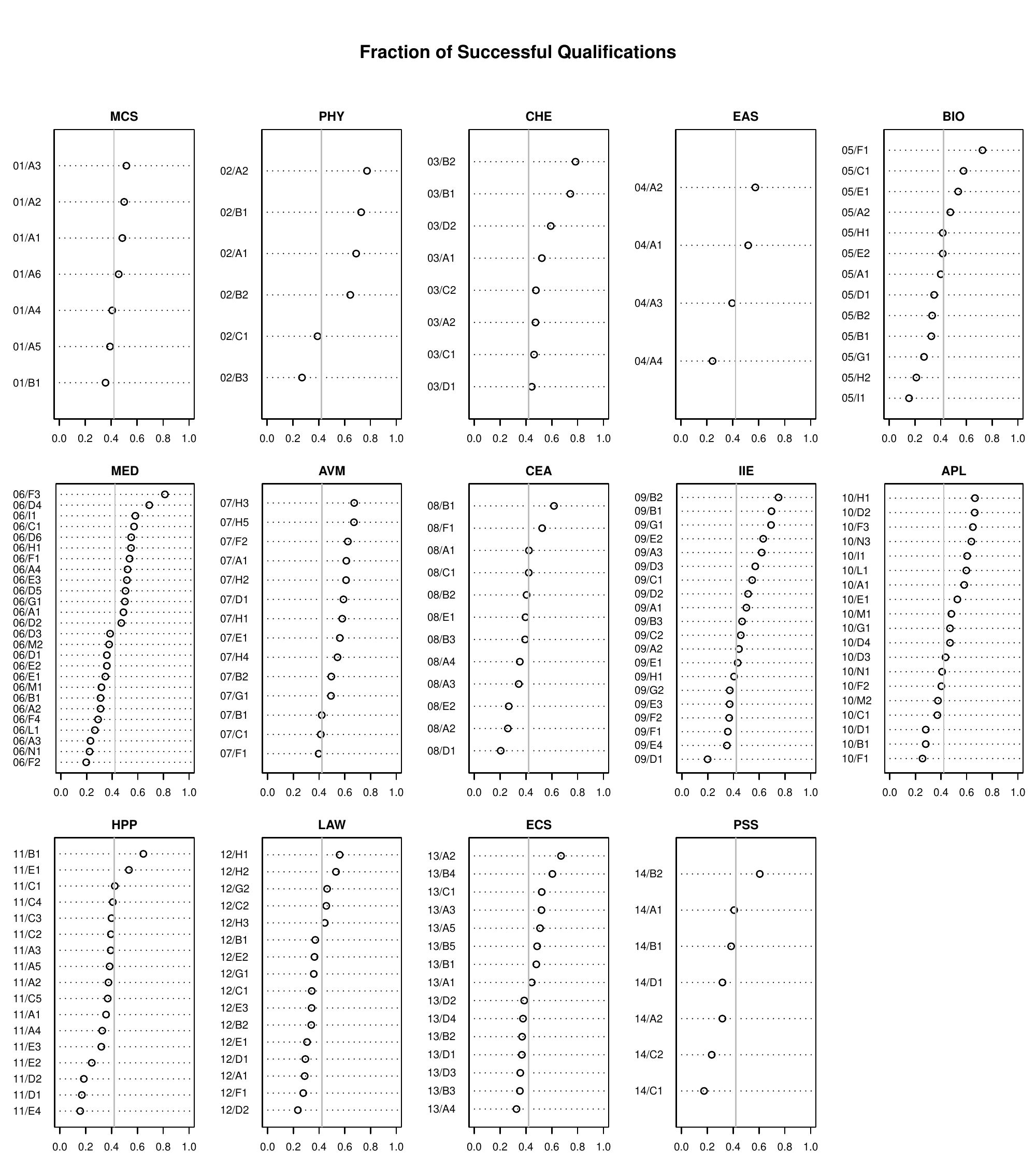}
\caption{Fraction of qualified applicants; the vertical lines denote
  the population median
  (0.421).}\label{fig:pr-qualified}
\end{figure}

Figure~\ref{fig:pr-qualified} shows the distribution of the fraction
of successful applications ($\PQ$) for each~\ac{SD}. The median
is~0.421, with inter-quartile
range~0.182. The five-number
summary is shown in Table~\ref{fn:frac-qualified}.

\begin{table}[ht]
  \centering%
  \caption{Fraction of qualified applicants}\label{fn:frac-qualified}
  \begin{fivenum}
    0.154 & 0.353 & 0.421 & 0.535 & 0.811\\
  \end{fivenum}
\end{table}

The distribution of $\PQ$ spans the interval
$[0.154,
  0.811]$, suggesting the presence
of several outliers. Indeed, it is interesting to compare the
five~\acp{SD} with lowest fraction of successful
qualification:\medskip

\begin{tabular}{ll}
0.154 & 05/I1--\emph{Genetics and microbiology}\\
0.157 & 11/E4--\emph{Clinical and dynamic psychology}\\
0.170 & 11/D1--\emph{Educational theories and history of educational theories}\\
0.175 & 14/C1--\emph{General and political sociology, sociology of law}\\
0.185 & 11/D2--\emph{Methodologies of teaching, special education and educational research}\\\end{tabular}\medskip

\noindent with the five~\acp{SD} with highest fraction of successful
qualification:\medskip

\begin{tabular}{ll}
0.743 & 03/B1--\emph{Principles of chemistry and inorganic systems}\\
0.750 & 09/B2--\emph{Industrial mechanical plants}\\
0.773 & 02/A2--\emph{Theoretical physics of fundamental interactions}\\
0.784 & 03/B2--\emph{Chemical basis of technology applications}\\
0.811 & 06/F3--\emph{Otorhinolaryngology and audiology}\\\end{tabular}\medskip

High variability also exists among disciplines of the same area. For
example, the values of $\PQ$ in area~\ac{BIO} ranges
from~0.154
(05/I1--\emph{Genetics and microbiology})
to~0.725
(05/F1--\emph{Experimental biology}).
Although statistical fluctuations could account for some of these
differences, it is hard to believe that the candidates of one
discipline are so much better (worse) than those of another. Each
scientific community has its own practices for evaluating researchers,
but these can still not justify the wide variations shown in
figure~\ref{fig:pr-qualified}. The identification of the root causes
of those differences is subject of ongoing research.

There is significant positive correlation between the fraction of
successful applications at the full and associate levels: $\rho(\PQF,
\PQA) = 0.77$ with 95\% \ac{CI}
$[0.70, 0.83]$. Therefore, whichever criteria
have been used for evaluating applicants, they have been applied
consistently to both roles.

\section{Medians}\label{sec:medians}

In this section we examine the medians that were used as thresholds of
the quantitative indicators of applicants. We address the following
questions:

\begin{itemize}
  \item How are medians distributed? (Section~\ref{sec:med-distribution})
  \item Are the medians for full professor qualification correlated
    with those for associate professor qualification?
    (Section~\ref{sec:med-correlation})
  \item Are the quantitative requirements for full professor
    qualification higher than those for associate professor
    qualification? (Section~\ref{sec:med-pareto})
\end{itemize}

In the following, $M_1.F$, $M_2.F$, $M_3.F$ denote the medians for
full professor qualification, and $M_1.A$, $M_2.A$, $M_3.A$ those for
associate professor qualification (the~\ac{SD} they refer to will be
irrelevant).

\subsection{Distributions}\label{sec:med-distribution}

\begin{figure}[t]

\centering%
\subfigure[Full professor\label{fig:medians-full}]{\includegraphics[width=\textwidth]{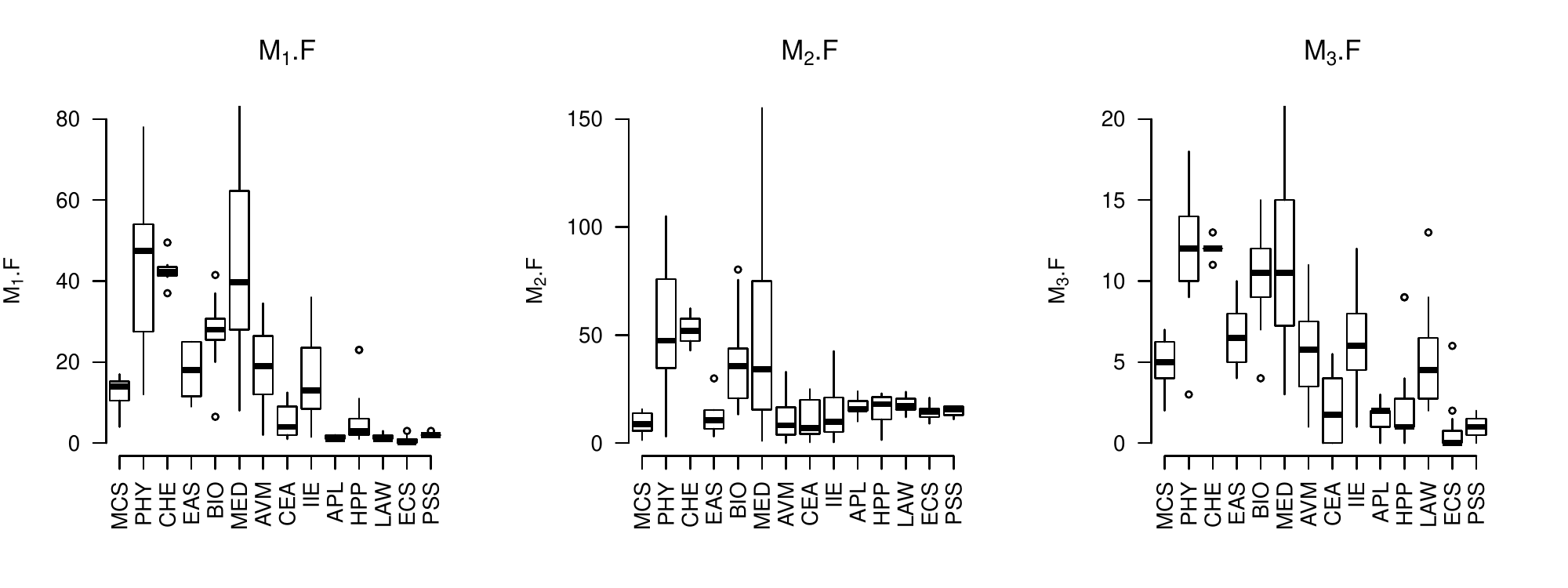}}\\
\subfigure[Associate professor\label{fig:medians-ass}]{\includegraphics[width=\textwidth]{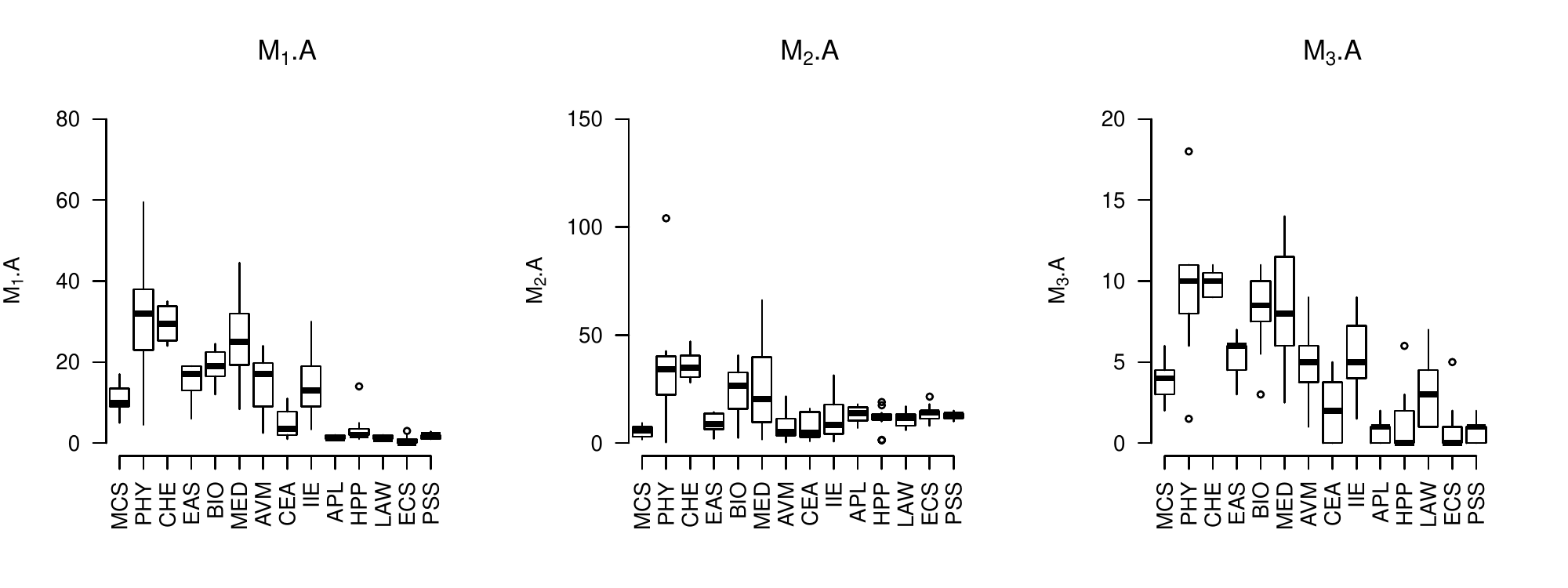}}
\caption{Distribution of medians for full professor
  qualification~\ref{fig:medians-full}, and associate professor
  qualification~\ref{fig:medians-ass}.}\label{fig:medians}
\end{figure}

Figure~\ref{fig:medians} shows the distributions of the values of
medians for full and associate professor qualification using box
plots~\citep{tukey1977exploratory}. Recall from
Section~\ref{sec:overview} that medians for bibliometric disciplines
refer to different types of indicators than those used for
non-bibliometric disciplines. In particular, non-bibliometric
indicators are based on paper-counting metrics, while bibliometric
indicators $B_2$ (normalized number of citations received) and $B_3$
(normalized $h$-index) are citation-based. This explains why medians
for non-bibliometric disciplines (\ac{APL}, \ac{HPP}, \ac{LAW},
\ac{ECS} and \ac{PSS}) are lower than those for bibliometric ones.

It is well known that impact metrics are not homogeneous across
scientific disciplines due to different publication and citation
patterns and practices~\citep{Hirsch2005}. Therefore, the large
variation of medians are originated from the different distributions
of bibliometric indicators~\citep{AlbarranCrOrRu11}. \ac{MED} is a
prominent example, with values for $M_1.F$, $M_2.F$ and $M_3.F$ that
span an order of magnitude.

An interesting observation is that some of the medians are
zero. Specifically, there are~30 disciplines
where one median for full professor qualification is zero,
and~6 disciplines where two medians for full
professor qualification are zero. Moreover, there
are~52 disciplines where one median for associate
professor qualification is zero, and~3 where two
medians for associate professor qualification are zero. Zero medians
are only present in some non-bibliometric disciplines; medians for
bibliometric indicators are all strictly positive.

Medians equal to zero are a strong hint that non-bibliometric
indicators, as defined in the~\ac{ASN}, may be not meaningful for all
disciplines they are applied to. To see why, remember that the medians
are based on quantitative indicators computed on the publications
submitted by tenured professors. Therefore, if a median is zero, then
the corresponding indicator is zero for at least half of tenured
professors. Such an indicator provides no or very little information
regarding the research profile of applicants, and should therefore be
revised.

\subsection{Correlations between Medians}\label{sec:med-correlation}

\begin{table}[t]
  \centering%
  \begin{tabular*}{.8\textwidth}{@{\extracolsep{\fill}}lllll}
    \toprule
        {\em Correlation of:} & \multicolumn{2}{l}{\em Bibliometric} & \multicolumn{2}{l}{\em Non-Bibliometric} \\
        \cmidrule{2-3}\cmidrule{4-5}
        & $\rho$ & 95\% CI & $\rho$ & 95\% CI \\
        \midrule
        $M_1.F$ vs $M_1.A$ & $0.91$ & $[0.87, 0.94]$ & $0.69$ & $[0.53, 0.80]$ \\
        $M_2.F$ vs $M_2.A$ & $0.97$ & $[0.95, 0.98]$ & $0.45$ & $[0.24, 0.62]$ \\
        $M_3.F$ vs $M_3.A$ & $0.95$ & $[0.93, 0.97]$ & $0.86$ & $[0.78, 0.92]$ \\
        \bottomrule
  \end{tabular*}
  \caption{Spearman's rank correlation $\rho(M_i.F, M_i.A)$ between
    the $i$-th median for full and associate professor qualification;
    95\% confidence intervals of the correlation coefficients are also
    reported.}\label{tab:cor-mediane}    
\end{table}

We expect that the medians for full professor qualification are
positively correlated with those for associate professor qualification
for the same discipline: if $M_i.F$ increases from one discipline to
another, we expect that $M_1.A$ increases as well. To test this
hypothesis, we compute the rank-order correlation coefficients
$\rho(M_i.F, M_i.A)$ between the $i$-th medians for full and associate
professor qualification for the same discipline ($i = 1, 2, 3$). We
consider bibliometric and non-bibliometric disciplines separately, to
see whether there are differences in the strength of the
association.

Table~\ref{tab:cor-mediane} shows that the medians for associate and
full professor qualification are indeed positively correlated for both
bibliometric and non-bibliometric disciplines. The correlation is
strong for bibliometric medians ($\rho > 0.90$); it is also high
between $M_1.F, M_1.A$ (normalized number of books) and $M_3.F, M_3.A$
(normalized number of papers published on top journals). On the other
hand, the correlation between $M_2.F$ and $M_2.A$ (normalized number
of journal papers) for non-bibliometric disciplines is weaker
($\rho(M_2.F, M_2.A) = 0.45$).

\subsection{Pareto Dominance Analysis}\label{sec:med-pareto}

It is reasonable to expect that the quantitative requirements for full
professor qualification are higher than those for associate professor
qualification for the same discipline, since a full professor must
demonstrate a stronger research profile and higher impact than an
associate professor. In the context of~\ac{ASN} this means that the
three medians for full professor qualification should be ``higher
than'' those for associate professor in the same discipline.

We can formalize this using the concept of \emph{Pareto dominance},
defined as follows: the $n$-dimensional real-valued vector $\mathbf{x}
= (x_1, \ldots x_n)$ Pareto-dominates $\mathbf{y} = (y_1, \ldots, y_n)$
(denoted with $\mathbf{x} \succ \mathbf{y}$) if the following
conditions hold:

\begin{enumerate}
  \item Every element of $\mathbf{x}$ is no lower than the
    corresponding element of $\mathbf{y}$: $x_i \geq y_i$ for each
    $i=1, \ldots, n$; and
  \item There exists at least one index $j \in \{1, \ldots, n\}$ for
    which the $j$-th element of $\mathbf{x}$ is strictly higher than
    the corresponding element of $\mathbf{y}$: $x_j > y_j$.
\end{enumerate}

We say that the quantitative requirements for full professor
qualification in some ~\ac{SD} are higher than those for associate
professor if and only if the medians $(M_1.F, M_2.F, M_3.F)$
Pareto-dominate $(M_1.A, M_2.A, M_3.A)$ for that
discipline. Surprisingly, this is not always true for the medians
defined for the~\ac{ASN}. First, there are several disciplines where
one of the medians for full professor qualification is lower than the
corresponding median for associate professor qualification, violating
the first condition above. Specifically, there
are~30
disciplines where $M_1.F < M_1.A$,
30
disciplines where $M_2.F < M_2.A$, and
8
where $M_3.F < M_3.A$; there are also disciplines where multiple violations
occurs, see the entries labeled ``O'', ``OO'', ``o'' and ``oo''
in~\ref{app:medians}.
Finally, there are~15~\acp{SD}
(10 bibliometric
and~5
non-bibliometric) where the medians for associate professor
qualification Pareto-dominate those for full professor qualification
(entries labeled ``*'' in the same table). This has the paradoxical
effect that applicants for a lower academic rank are required by
the~\ac{ASN} to satisfy higher quantitative standards than applicants
for a higher rank.

The situations above may have happened for several reasons. The most
obvious explanation could be that older generations of scholars are
less productive than younger researchers; however, this explanation
seems refuted by a recent study~\citep{AbramoDADi11}. Also, errors or
omissions in the publication list that was used to compute the medians
may have introduced distortions in the raw data (that is unfortunately
not publicly available and therefore can not be audited). Finally, in
some disciplines full professor may be unable to dedicate much time to
research, due to other administrative tasks or higher teaching load
than associate professors. Whatever the reason, the existence of
Pareto violations suggests that the thresholds for qualification
should be defined with a different, more robust mechanism.

\section{Analysis of the Quantitative Indicators of Applicants}\label{sec:bib-indicators}

In this section we analyze the values of the quantitative indicators
used in the evaluation of applicants. In the following we denote
applicants whose indicators exceed two medians (one, for
non-bibliometric disciplines) as ``over-median'', and those who do not
satisfy this requirement as ``under-median''.

We address the following questions:
\begin{itemize}
  \item Are the quantitative indicators pairwise correlated?
    (Section~\ref{sec:ind-correlation})
  \item Were over-median applicants more likely to get qualification
    than under-median ones? (Section~\ref{sec:cond-qual-prob})
  \item What are the minimum values of each quantitative
    indicator below which qualification is not granted?
    (Section~\ref{sec:min-values})
  \item Do the~\ac{ASN} results preserve the Pareto dominance relation
    between applicants? (Section~\ref{sec:pareto-dom-an})
\end{itemize}

\subsection{Correlation between Quantitative Indicators}\label{sec:ind-correlation}

\begin{figure}[t]
  \begin{center}
    \subfigure[Full professor, bibliometric\label{fig:pairs-full-bib}]{\includegraphics[trim=30 30 30 30,clip,width=.3\textwidth]{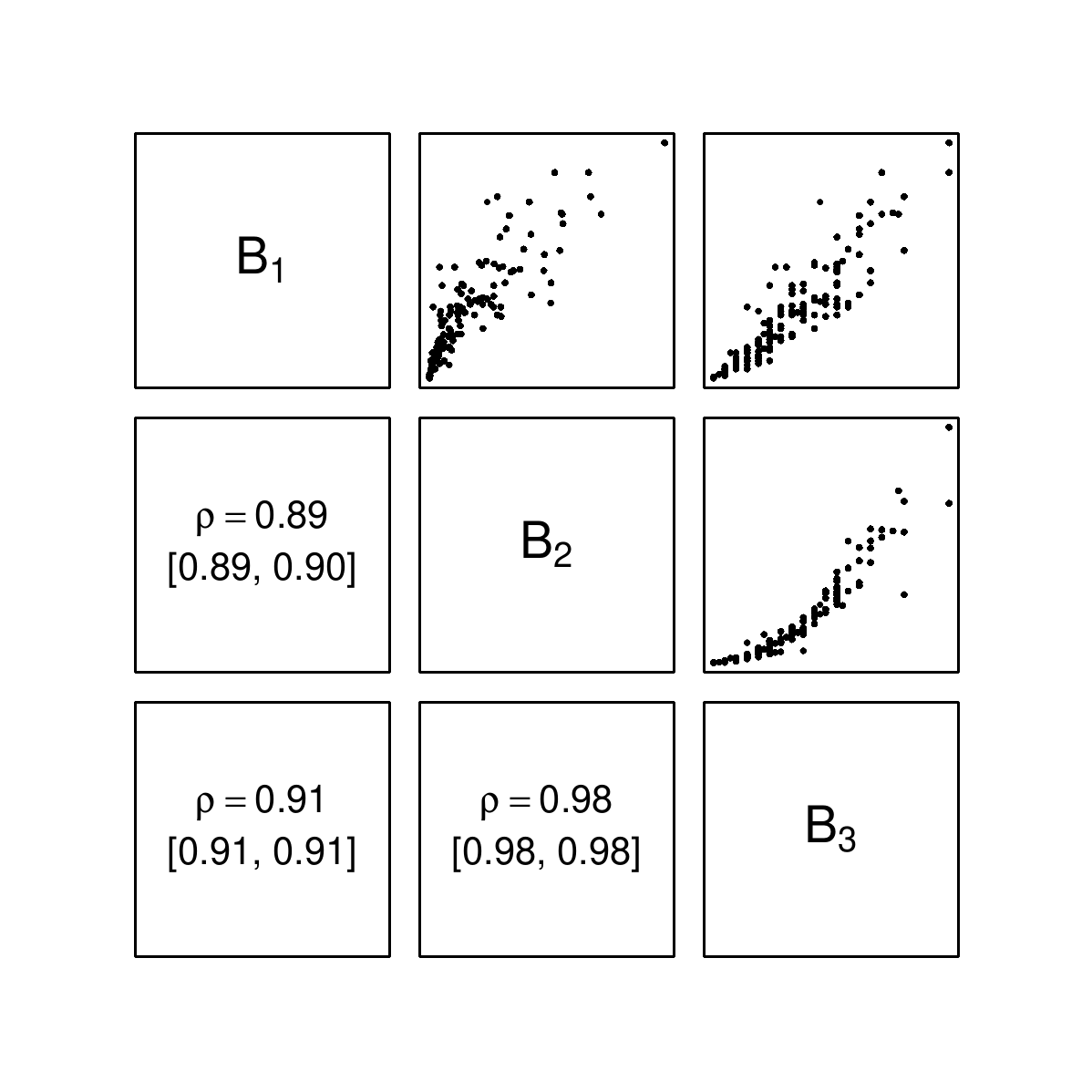}} \qquad
    \subfigure[Full professor, non-bibliometric\label{fig:pairs-ass-bib}]{\includegraphics[trim=30 30 30 30,clip,width=.3\textwidth]{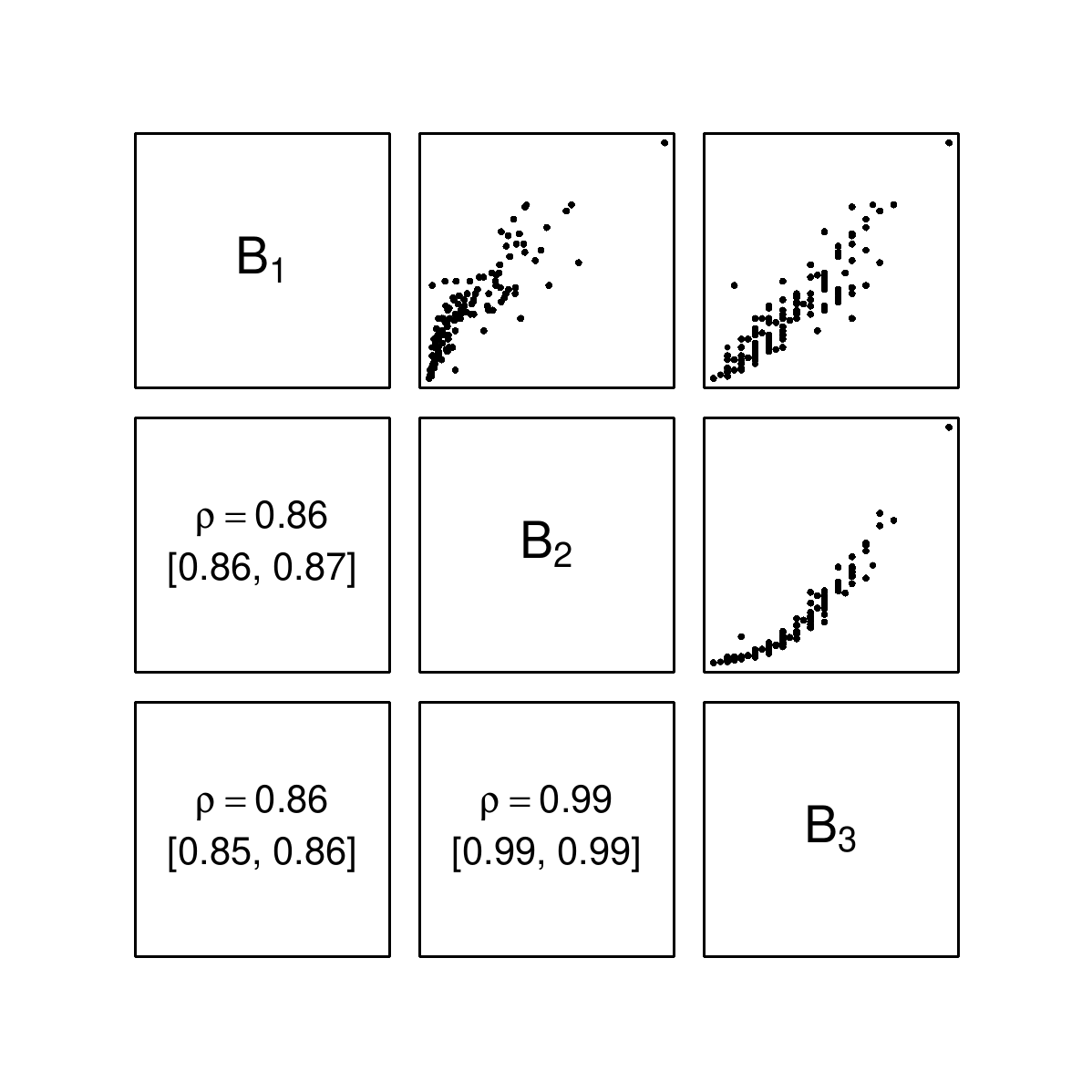}}\\
    \subfigure[Associate professor, bibliometric\label{fig:pairs-full-nbib}]{\includegraphics[trim=30 30 30 30,clip,width=.3\textwidth]{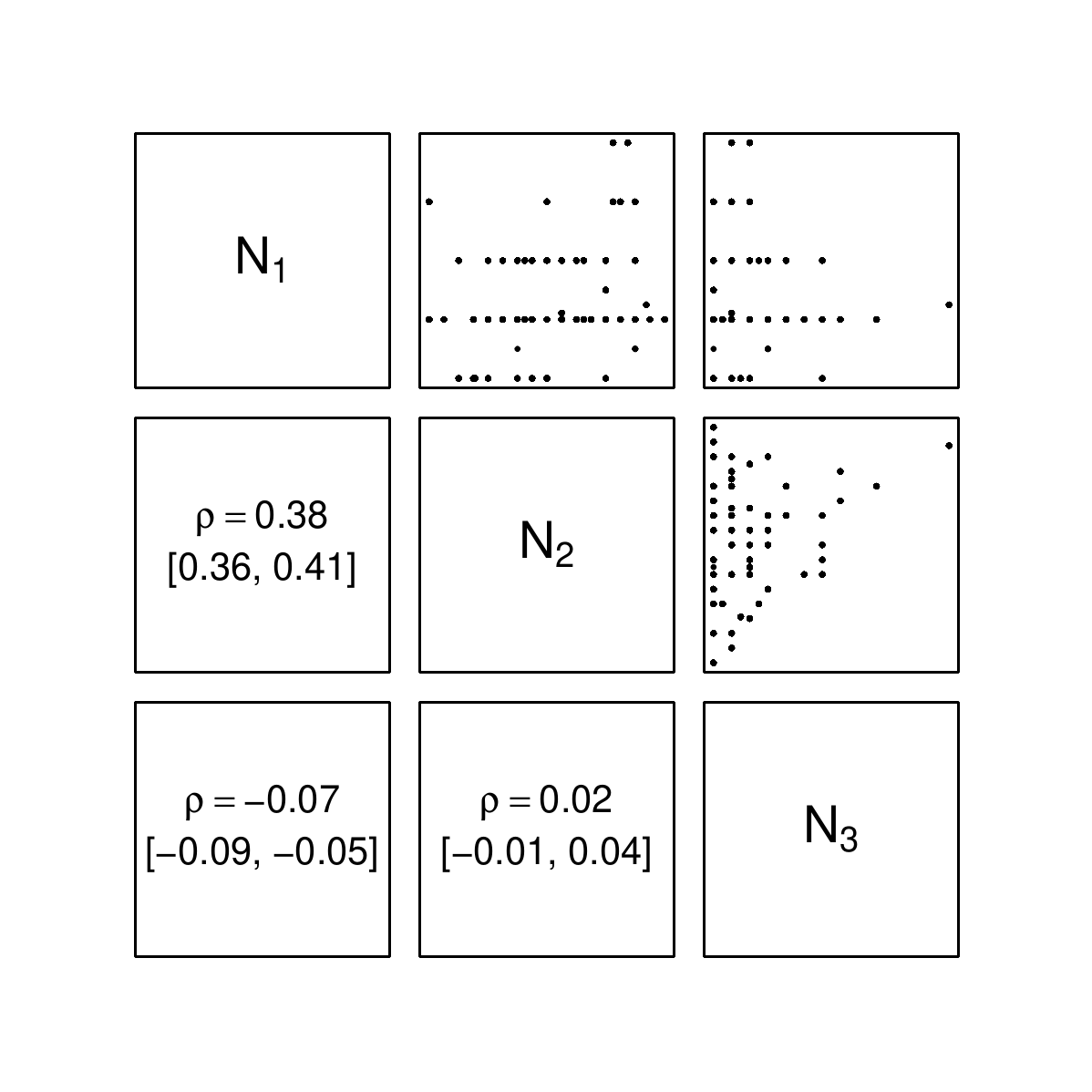}} \qquad
    \subfigure[Associate professor, non-bibliometric\label{fig:pairs-ass-nbib}]{\includegraphics[trim=30 30 30 30,clip,width=.3\textwidth]{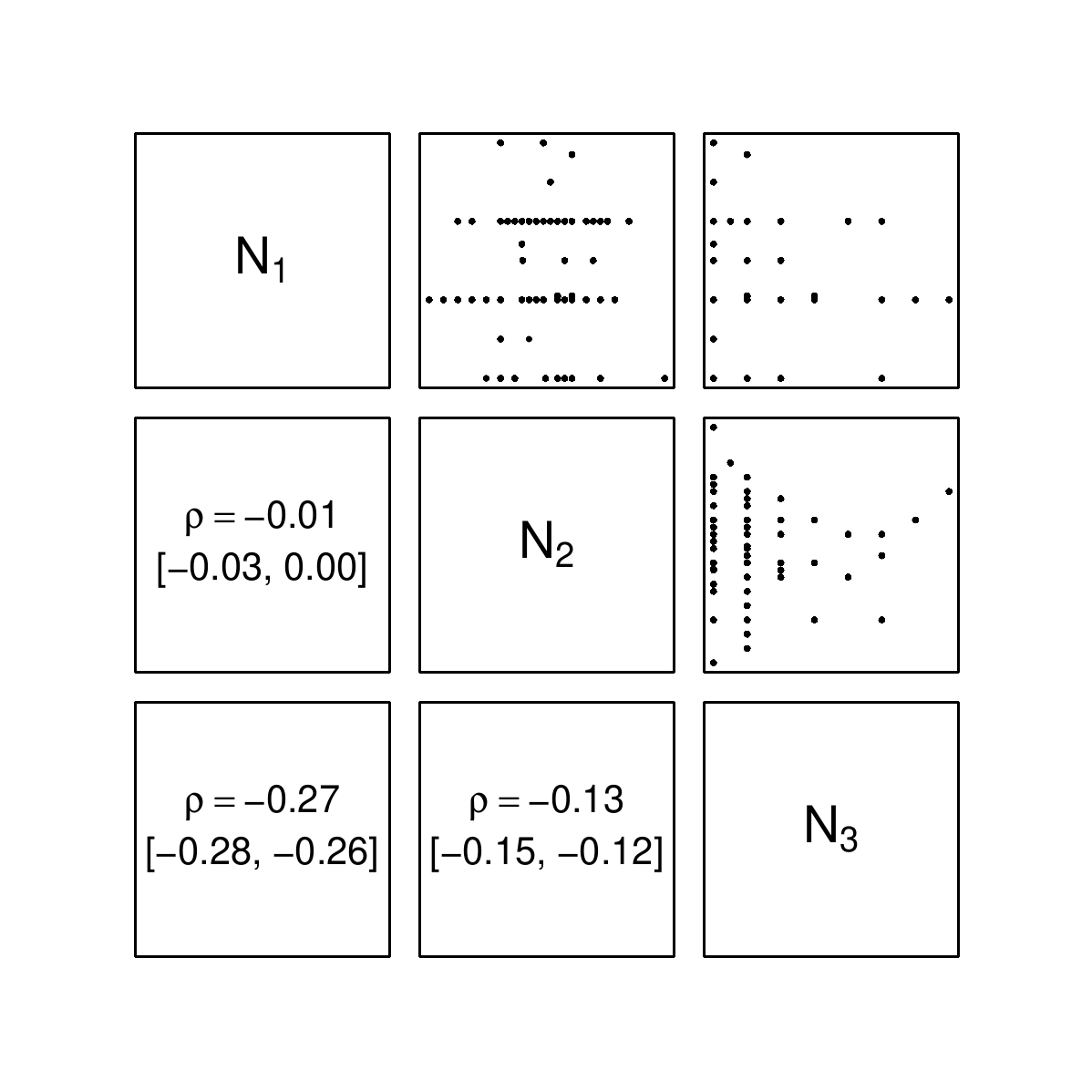}} 
  \end{center}
\caption{Each scatter plot displays the correlation between pairs of
  quantitative indicators for the same role; the correlation
  coefficients are shown with 95\% CIs. Each data point corresponds to
  an applicant.}\label{fig:pairs-indicators}
\end{figure}

We start our analysis by testing whether the values of different
quantitative indicators among applicants for the same role are
correlated. It seems reasonable to expect that applicants with higher
values of one indicator have higher values of other indicators as
well, therefore we expect positive correlation. To verify this
hypothesis we compute the Spearman's rank correlation coefficient
between the $i$-th and $j$-th indicators, $i \neq j$ for each of the
following subsets of applications:

\begin{itemize}
\item Applications for full professor qualification in bibliometric
  disciplines ($\rho(B_i.F, B_j.F)$, $i \neq j$);
\item Applications for associate professor qualification in
  bibliometric disciplines ($\rho(B_i.A, B_j.A)$, $i \neq j$);
\item Applications for full professor qualification in non-bibliometric
  disciplines ($\rho(N_i.F, N_j.F)$, $i \neq j$);
\item Applications for associate professor qualification in
  non-bibliometric disciplines qualification ($\rho(N_i.A, N_j.A)$, $i
  \neq j$).
\end{itemize}

The results are shown in the four scatter plot matrices~\citep{eda} of
Figure~\ref{fig:pairs-indicators}. Each matrix refers to one of the
four possible combinations of (full, associate) qualification for
(bibliometric, non-bibliometric) disciplines. Cell $(i, j)$ in the
upper triangular part shows the scatter plot of $B_i$ (resp. $N_i$)
versus $B_j$ (resp. $N_j$); every point represents one application.
Cell $(j, i)$ in the lower triangular part shows the corresponding
rank correlation coefficient $\rho$ and the $p$-value of the null
hypothesis of no correlation.

Figures~\ref{fig:pairs-full-bib} and~\ref{fig:pairs-ass-bib} show that
quantitative indicators for both full and associate professor
applicants in bibliometric disciplines are strongly pairwise
correlated. This confirms our expectation, and suggests that the
indicators defined for bibliometric disciplines may indeed reflect
different aspects of a common quantitative profile of each
candidate. On the other hand, Figures~\ref{fig:pairs-full-nbib}
and~\ref{fig:pairs-ass-nbib} show that quantitative indicators for
non-bibliometric disciplines are not pairwise correlated, and
therefore it is not clear what they measure. This is another strong
call for better understanding whether the indicators defined for the
social sciences and humanities are a meaningful way for assessing
researchers in those research areas.

\subsection{Conditional Qualification Probabilities}\label{sec:cond-qual-prob}

\begin{figure}[t]
\centering\includegraphics[width=\textwidth]{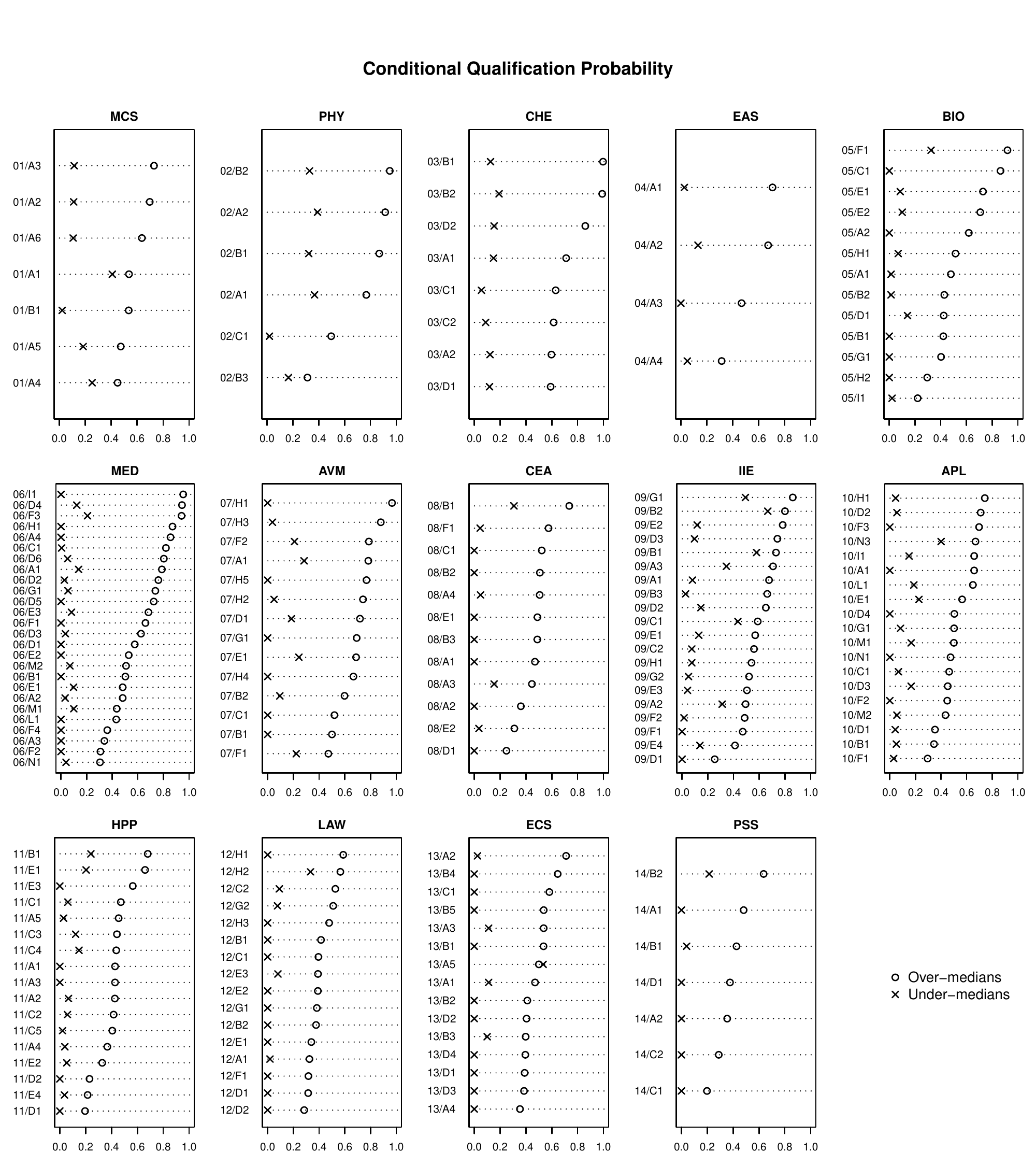}
\caption{Fraction of qualified over- and under-median applicants; data sorted in decreasing value for over-median applicants.}\label{fig:cond-qual-prob}
\end{figure}

Are over-median applicants more likely to get qualification than
under-median ones? To answer this question we compute the fractions
$\PQO$, $\PQU$ of over- and under-median applicants that received
qualification. These quantities can be seen as conditional
qualification probabilities defined as:

\begin{align}
\PQO & = \frac{\text{\# of qualified over-median applicants}}{\text{\# of over-median applicants}} \medskip \\ 
\PQU & = \frac{\text{\# of qualified under-median applicants}}{\text{\# of under-median applicants}}
\end{align}

At the global level, 77.2\% of the applications are
over-median (80.8\% of applications for full
professor qualification, and 75.6\% of those for
associate professor qualification).  52.8\% of
over-median applicants got qualification
(51.5\% at the full professor level,
53.4\% at the associate professor level),
compared with 9.2\% of under-median applicants
(8.4\% at the full professor level, and
9.4\% at the associate professor level).

Figure~\ref{fig:cond-qual-prob} shows the values of $\PQO$ and $\PQU$
for each~\ac{SD}. The fractions of qualified under-median applicants
are in general much lower than the fractions of qualified over-median
applicants (in 69 disciplines no
under-median applicant got qualification). There is a single
exception, 13/A5--\emph{Econometrics},
for which
$\PQO=0.500$
and
$\PQU=0.533$.
Of course, it is not possible to claim any causal relationship between
exceeding medians and getting qualification. In fact, it is equally
possible that (i) examination committees were somewhat biased towards
granting qualification to over-median applicants (the~\ac{ASN} rules
encourage this), or that (ii) over-median applicants have higher
quantitative indicators because they are intrinsically ``better'', and
therefore more likely to qualify anyway.

\begin{table}[ht]
  \centering%
  \caption{Fraction of qualified over-median applicants}\label{fn:frac-qual-over}
  \begin{fivenum}
    0.195 & 0.419 & 0.509 & 0.684 & 0.997 \\  
  \end{fivenum}
\end{table}

\begin{table}[ht]
  \centering%
  \caption{Fraction of qualified under-median applicants}\label{fn:frac-qual-under}
  \begin{fivenum}
    0 & 0 & 0.041 & 0.123 & 0.667 \\  
  \end{fivenum}
\end{table}

Looking at the five-number summaries for $\PQO$
(Table~\ref{fn:frac-qual-over}) and $\PQU$
(Table~\ref{fn:frac-qual-under}) we observe that there are~\acp{SD}
with either very low or very high fractions of over-median applicants
that got qualification. The median of $\PQO$ is slightly more than
$0.5$, meaning that 50\% of the examination committees denied
qualification to more than half of over-median applicants. Therefore,
exceeding the medians is loosely correlated with getting qualification
in halt of the~\acp{SD}.

\begin{table}[t]
\begin{center}
\begin{tabular*}{\textwidth}{@{\extracolsep{\fill}}llllllll}
  \toprule
  {\em Fraction of qualified:} & \multicolumn{3}{l}{\em Full professor} & \multicolumn{3}{l}{\em Associate professor} \\
  \cmidrule{2-4}\cmidrule{5-7} 
  & B & N & 95\% CI $(\textrm{B}-\textrm{N})$ & B & N & 95\% CI $(\textrm{B}-\textrm{N})$ \\
\midrule
 over-median applicants & 0.568 & 0.440 & [0.11, 0.14] & 0.606 & 0.452 & [0.14, 0.17] \\ 
  under-median applicants & 0.087 & 0.070 & [-0.01, 0.04] & 0.109 & 0.040 & [0.06, 0.08] \\ \bottomrule
\end{tabular*}
\end{center}
\caption{Fraction of qualified over- and under-median applicants in
  bibliometric (B) and non-bibliometric (N) disciplines. Columns
  labeled $(\mathrm{B}-\mathrm{N})$ shows the 95\% \ac{CI} of the
  difference $(\mathrm{B}-\mathrm{N})$; positive values indicate that
  the proportion of qualified over- (resp. under-) median applicants
  in bibliometric disciplines is higher than in non-bibliometric
  ones.}\label{tab:prob-qual}
\end{table}

To test whether there are differences across bibliometric and
non-bibliometric disciplines and across roles, we compute the fraction
of qualified over- and under-median applicants for each of the
following disjoint sets:

\begin{itemize}
\item Applicants for full professor qualification on bibliometric
  disciplines;
\item Applicants for associate professor qualification on bibliometric
  disciplines;
\item Applicants for full professor qualification on non-bibliometric
  disciplines;
\item Applicants for associate professor qualification on
  non-bibliometric disciplines.
\end{itemize}

The results are reported in Table~\ref{tab:prob-qual}. Within the four
classes above we observe the same general pattern that we have seen
above for individual disciplines: over-median applicants were on
average more likely to get qualification than under-median
ones. Additionally, over-median applicants were more likely to get
qualification in bibliometric disciplines rather than non-bibliometric
ones; the differences are non-negligible and statistically
significant. The reason of these differences is yet to be identified.

\begin{figure}[t]
  \begin{center}
    \subfigure[\label{fig:cor-conditional-ex-bib}]{\includegraphics[width=.4\textwidth]{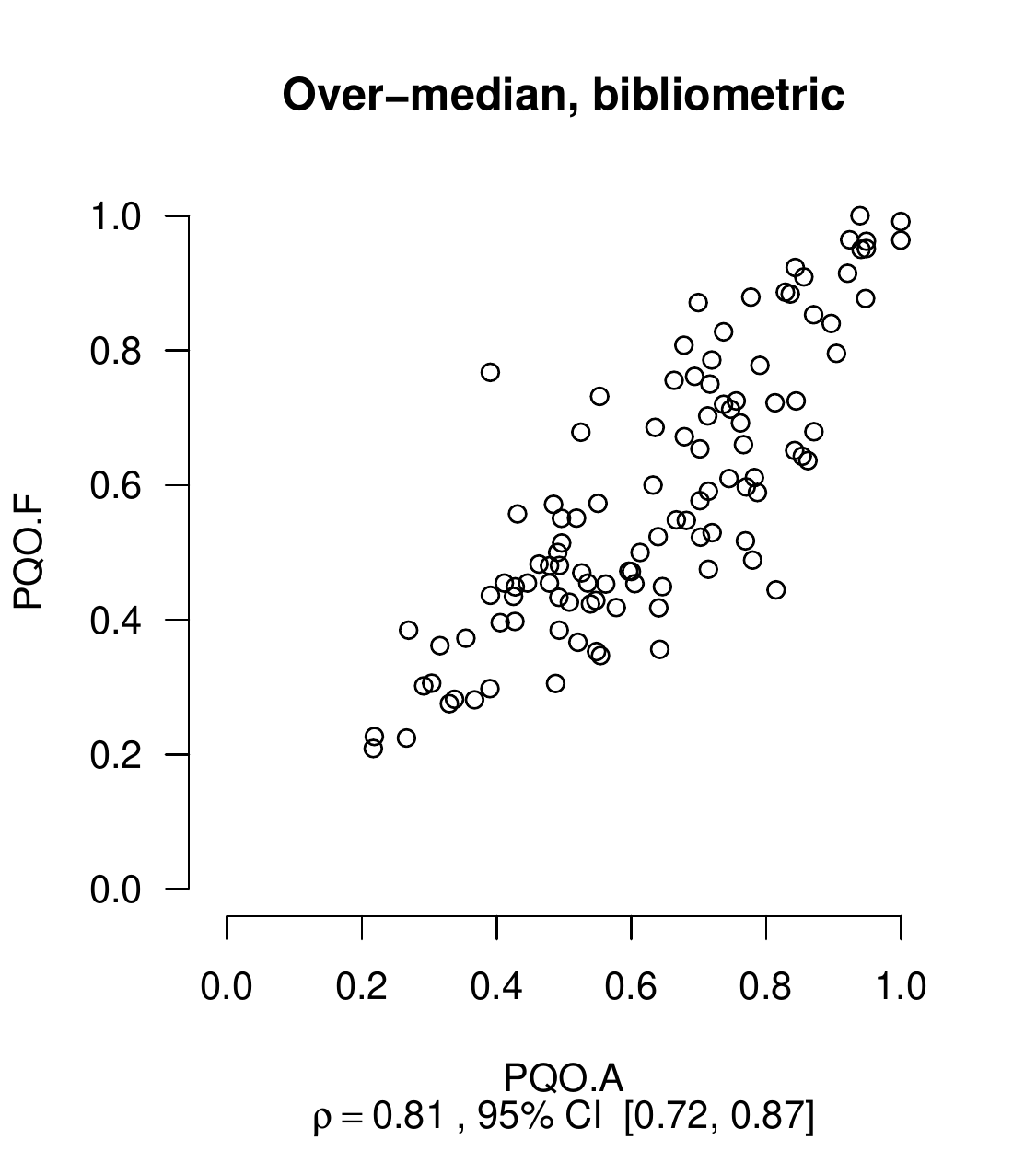}}\qquad
    \subfigure[\label{fig:cor-conditional-ex-nbib}]{\includegraphics[width=.4\textwidth]{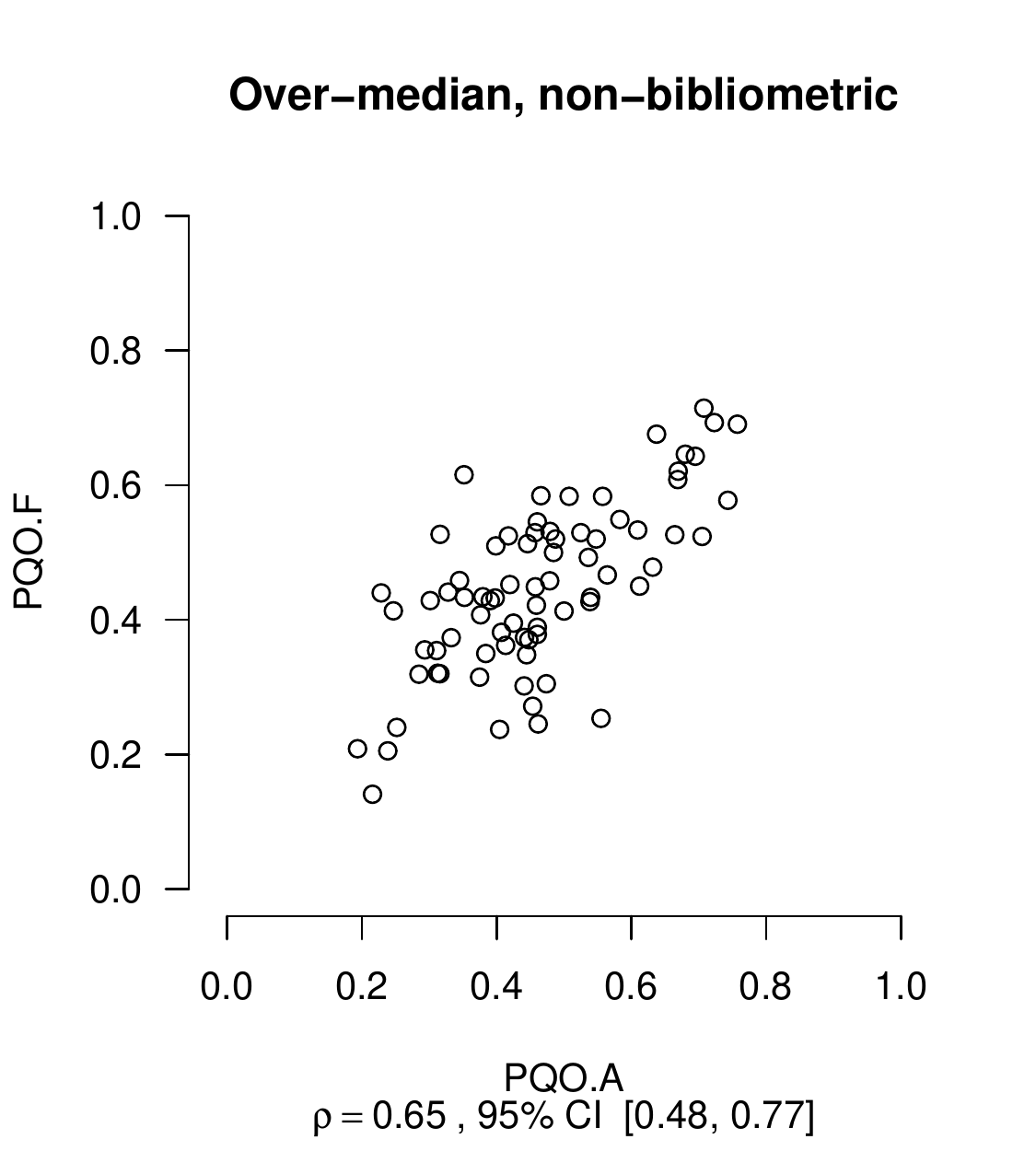}}\\
    \subfigure[\label{fig:cor-conditional-nex-bib}]{\includegraphics[width=.4\textwidth]{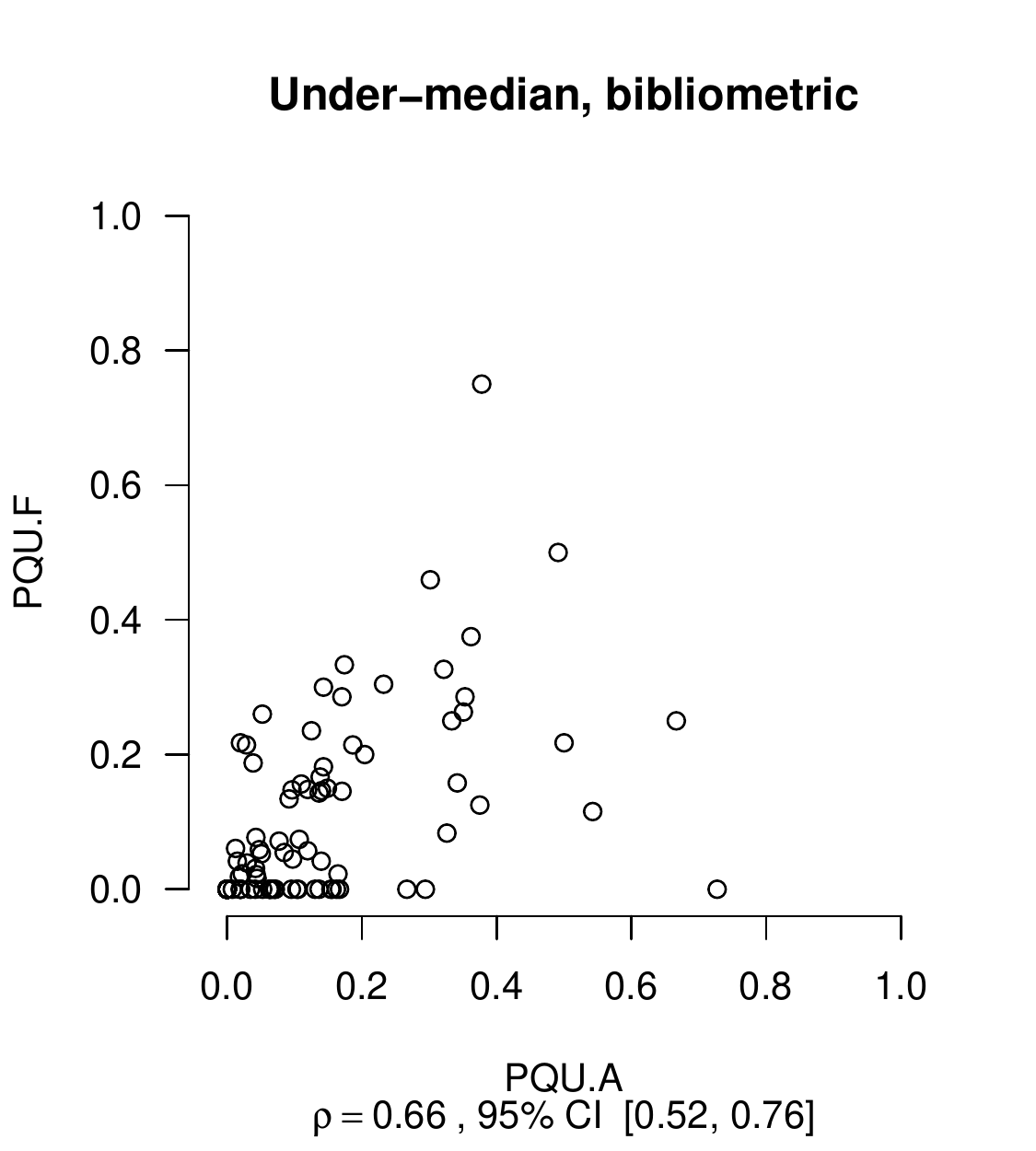}}\qquad
    \subfigure[\label{fig:cor-conditional-nex-nbib}]{\includegraphics[width=.4\textwidth]{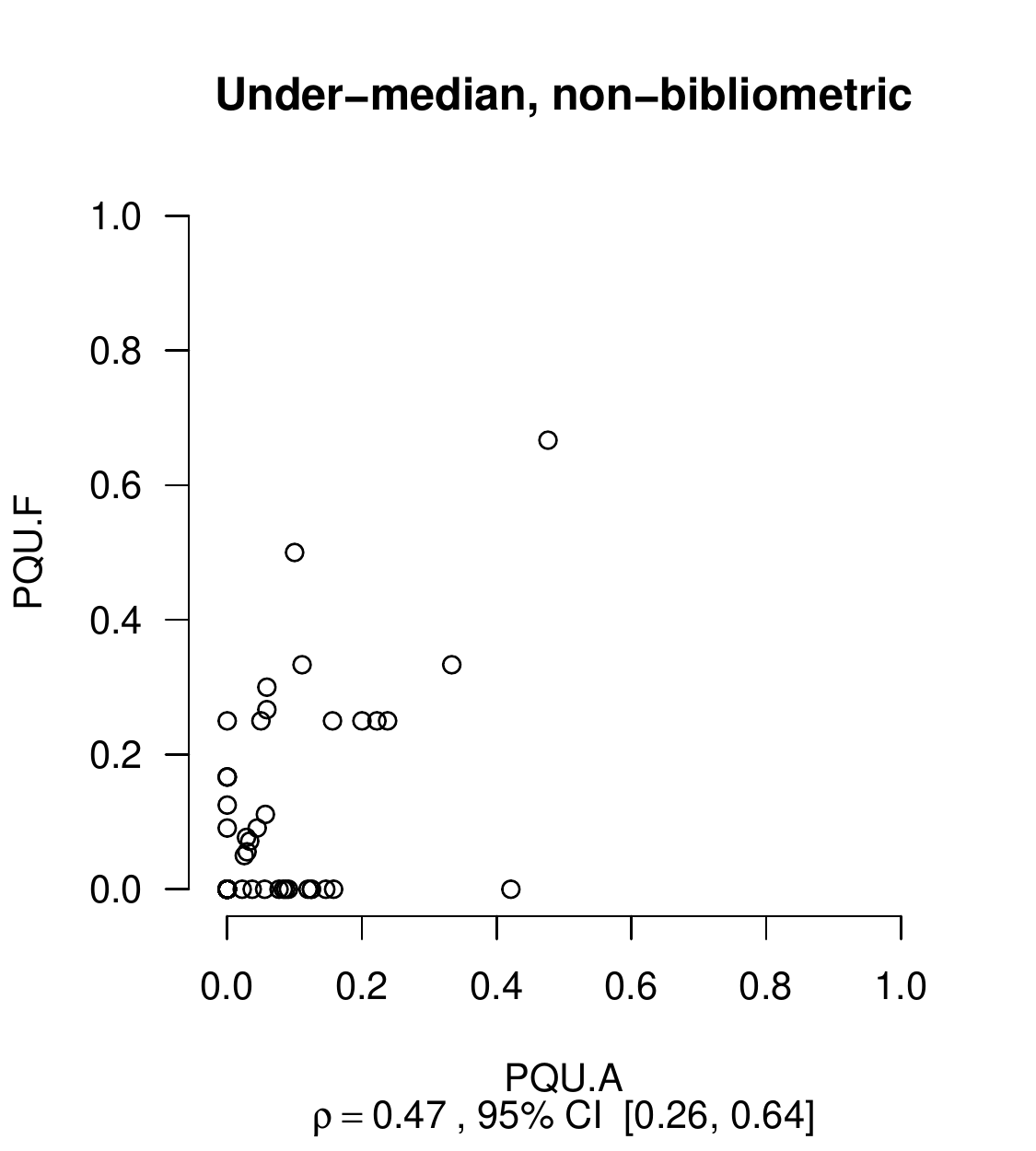}}
  \end{center}
  \caption{Qualification probabilities for over-median
    (Fig.~\ref{fig:cor-conditional-ex-bib}
    and~\ref{fig:cor-conditional-ex-nbib}) and under-median
    (Fig.~\ref{fig:cor-conditional-nex-bib}
    and~\ref{fig:cor-conditional-nex-nbib}) applicants at the the full
    versus associate professor levels. Each point represents
    a~\ac{SD}}\label{fig:cor-conditional}
\end{figure}

Finally, we compute the strength of the association between the
conditional qualification probabilities for full and associate
professor qualification. Figures~\ref{fig:cor-conditional-ex-bib}
and~\ref{fig:cor-conditional-nex-bib} show the correlation between the
fraction of over-median qualified applicants for the full ($\PQOF$)
and associate ($\PQOA$) levels, for bibliometric and non-bibliometric
disciplines, respectively; Figures~\ref{fig:cor-conditional-ex-nbib}
and~\ref{fig:cor-conditional-nex-nbib} show the correlation between
$\PQUF$ and $\PQUA$ for bibliometric and non-bibliometric disciplines.
Correlation is high for over-median applicants for bibliometric
disciplines ($\rho=0.81$), and still significant for
non-bibliometric ones ($\rho=0.65$). Positive
correlation suggests that evaluation criteria have been applied
consistently for both roles.

The qualification probabilities for under-median applicants are
positively correlated between full and associate roles, although with
lower strength both for bibliometric disciplines
($\rho=0.66$) and non-bibliometric ones
($\rho=0.47$). This suggests that the decisions to
grant or deny qualification to under-median applicants were taken on a
case-by-case basis.

\subsection{Minimum Values of the Indicators for Qualified Applicants}\label{sec:min-values}

In this section we address the following question: are there minimum
values for each quantitative indicator below which qualification has
not been given? From the data on Table~\ref{tab:prob-qual} we already
know that there are qualified under-median applicants, so we expect
that many of those minimum values are below the medians.

\begin{figure}[t]

\centering%
\subfigure[Full professor\label{fig:min-full}]{\includegraphics[width=\textwidth]{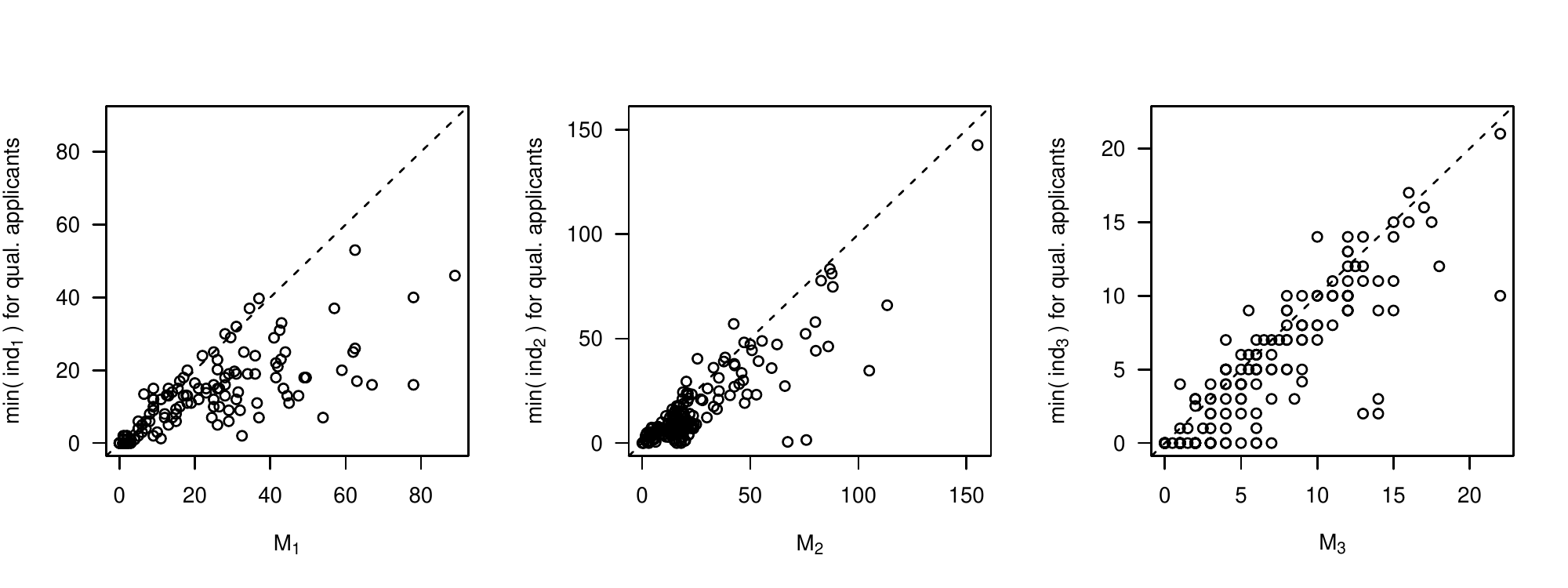}}\\
\subfigure[Associate professor\label{fig:min-ass}]{\includegraphics[width=\textwidth]{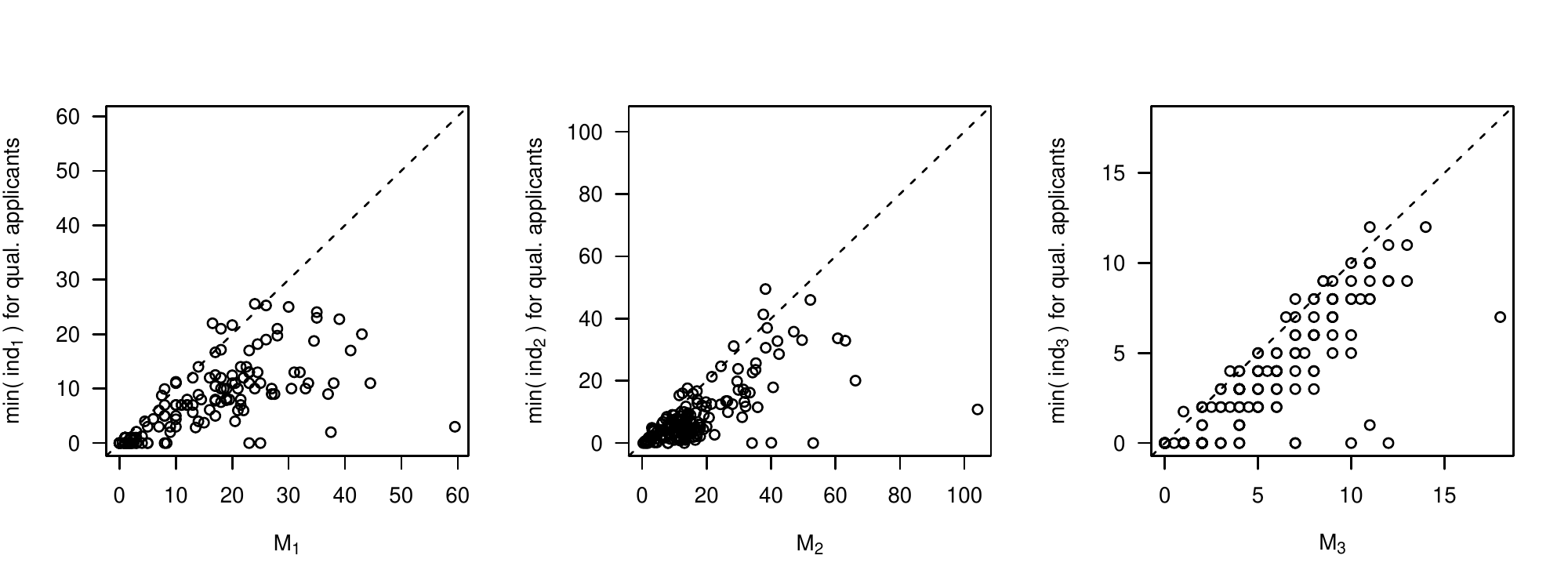}}
\caption{Minimum value of the quantitative indicators for
  full~\ref{fig:min-full} and associate professor
  qualification~\ref{fig:min-ass}}\label{fig:min-qual}
\end{figure}

Each point in Figure~\ref{fig:min-qual} represents a~\ac{SD}; the $x$
coordinate is the value of one of the medians, while the $y$ coordinate
is the minimum value of the corresponding indicator among all
successfully qualified applicants. Points located below the dashed
lines denote~\acp{SD} where the minimum value among qualified
applicants is lower than the median. Points above the lines
denote~\acp{SD} where the minimum value among qualified applicants is
higher than the median.

Figure~\ref{fig:min-qual} shows the values of each indicator that the
examination committees considered the ``absolute minimum'' for
granting qualification, i.e., no applicant with a lower values got
qualified.

\begin{table}[t]
  \centering
  \begin{tabular}{llll}
    \toprule
    Role & $\#\{ \min(\mathit{ind}_1) > M_1\}$ & $\#\{ \min(\mathit{ind}_2) > M_2\}$ & $\#\{\min(\mathit{ind}_3) > M_3\}$ \\
    \midrule
    Full Professor &
    $20$ & 
    $33$ & 
    $27$ \\ 

    Associate Professor & 
    $8$ & 
    $18$ & 
    $7$ \\ 
    \bottomrule
  \end{tabular}
  \caption{Number of~\acp{SD} for which qualification has been granted
    only to applicants whose value of $\mathit{ind}_i$ is strictly
    higher than the corresponding median $M_i$}\label{tab:min-qual}
\end{table}

We observe that, for most disciplines, the minimum values of the
quantitative indicators of successful applicants are below the
medians. However, there are disciplines where the minimum value of a
quantitative indicator across qualified applicants is higher than the
median. In Table~\ref{tab:min-qual} we report, for each indicator
$\mathit{ind}_i$ and role, the number of disciplines for which the
minimum value of $\mathit{ind}_i$ for qualified applicants is strictly
higher than the corresponding median $M_i$. We observe that this
happened more frequently for full professor qualification, suggesting
that examination committees enforced stronger quantitative
requirements than those required by the~\ac{ASN} rules.

\subsection{Pareto Dominance Analysis}\label{sec:pareto-dom-an}

The Pareto dominance relation introduced in
Section~\ref{sec:med-pareto} can be used to define a partial order
among applicants. Let us consider two researchers, Alice and Bob,
applying for the same role in the same~\ac{SD}. Suppose that the
quantitative indicators for Alice are $I_\text{Alice} = (11, 8, 15)$
and those for Bob are $I_\text{Bob} = (10, 8, 13)$. Then, Alice
Pareto-dominates Bob since $I_\text{Alice} \succ I_\text{Bob}$; in
other words Alice is quantitatively ``no worse than'' Bob with respect
to all indicators, and strictly ``better'' in two of them. Thus, if
Bob gets the qualification we expect that also Alice does.

If Bob gets the qualification but Alice does not, we have a violation
of Pareto dominance. There could be many valid reasons for this to
happen: for example, Alice could have applied to a~\ac{SD} unrelated
to her research field, or she could have failed to meet the minimum
qualitative requirements for the role applied for. Therefore, Pareto
violations do not automatically indicate a problem, but nevertheless
represent anomalies that require further investigation.

To study the frequency of Pareto violations in each discipline, we
define a metric called~\ac{PVR} as follows. Let $\mathit{AP}$ be the
set of applicants to a given~\ac{SD} and role. Let $p, q$ be two
applicants in $\mathit{AP}$ such that $p$ Pareto-dominates $q$. Let
$I_p$ and $I_q$ the vectors of the quantitative indicators of $p$ and
$q$, respectively. We have a Pareto violation if $p$ is not qualified
but $q$ is; in all other cases there is no violation, as summarized in
Table~\ref{tab:violations}. 

\begin{table}[ht]
  \centering%
  \begin{tabular}{lll}
    \toprule
    $p$ & $q$ & {\em Pareto-violation?} \\
    \midrule
    Qualified & Qualified & No \\
    Qualified & Not qualified & No \\
    Not qualified & Qualified & Yes \\
    Not qualified & Not qualified & No \\
    \bottomrule
  \end{tabular}
  \caption{Condition for Pareto-violation between two applicants $p$
    and $q$ where the quantitative indicators of $p$ Pareto-dominate
    those of $q$.}\label{tab:violations}
\end{table}

\noindent We can therefore define~\ac{PVR} as:

\begin{equation}
\mathit{PVR} = \frac{\#\{ (p, q) \in \mathit{AP} \times \mathit{AP}\ s.t.\ I_p \succ I_q \wedge \text{$p$ not qualified} \wedge \text{$p$ qualified}\}}{\#\{ (p, q) \in \mathit{AP} \times \mathit{AP}\ s.t.\ I_p \succ I_q\}}
\end{equation}

\noindent where $\#\{ X\}$ is the cardinality (number of elements) of
set $X$.  By definition, $0 \leq \PVR \leq 1$, where $\PVR = 0$ if no
qualified applicant is Pareto-dominated by a not qualified one, while
$\PVR = 1$ if for every pairs of applicants $p, q$ where $p$
Pareto-dominates $q$ we have that $p$ is not qualified while $q$
is. Therefore, higher~\ac{PVR} values indicate anomalous situation
that should be investigated.

\begin{figure}[t]
\centering\includegraphics[width=\textwidth]{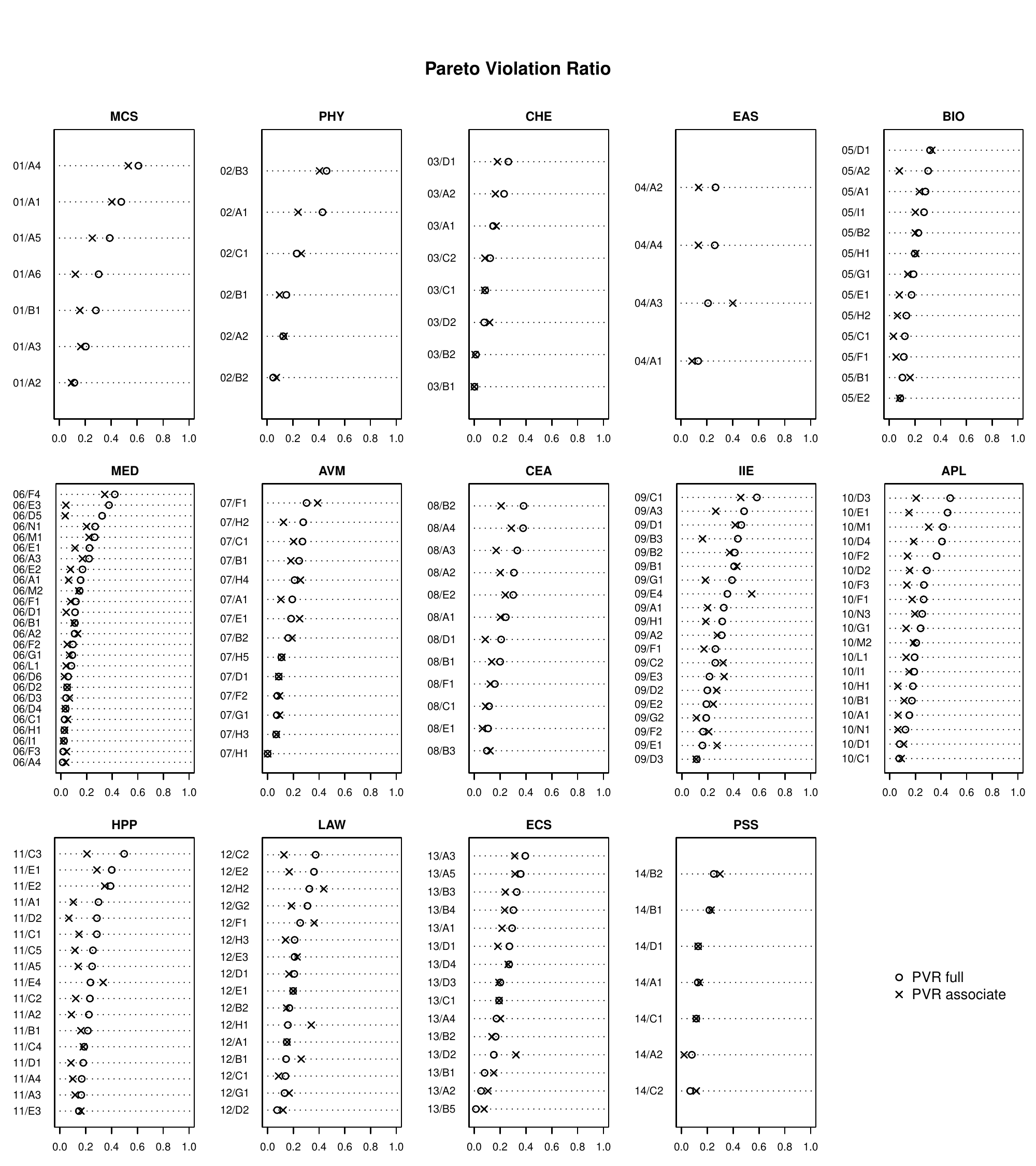}
\caption{Pareto Violation Ratio for full and associate professor
  qualification; data sorted by $\PVRF$.}\label{fig:pvr}
\end{figure}

Figure~\ref{fig:pvr} shows the~\ac{PVR} values among full ($\PVRF$)
and associate ($\PVRA$) professor qualification for each
discipline. In most cases, the~\ac{PVR} for full professor applicants
is higher than that for associate professor applicants (there are
exactly~57
\acp{SD} where $\PVRF < \PVRA$).
The values of~\ac{PVR} for the full and associate roles are positively
correlated, although the strength of the correlation is higher for
bibliometric disciplines ($\rho=0.77$) than
non-bibliometric ones ($\rho=0.51$). The
five-number summaries indicate that the values of~$\PVRF$
(Table~\ref{fn:pvr-full}) and~$\PVRA$ (Table~\ref{fn:pvr-ass}) are
quite small: the maximum of $\PVRF$ is $0.61$,
while the maximum of $\PVRA$ is $0.542$. However,
the relative width of the interval from the third quartile to the
maximum is large for both $\PVRF$ and $\PVRA$, denoting the presence
of several outliers.

\begin{table}[ht]
  \centering%
  \caption{PVR for Full Professor Qualifications ($\PVRF$)}\label{fn:pvr-full}
  \begin{fivenum}
    0 & 0.123 & 0.203 & 0.29 & 0.61 \\
  \end{fivenum}
\end{table}

\begin{table}[ht]
  \centering%
  \caption{PVR for Associate Professor Qualifications ($\PVRA$)}\label{fn:pvr-ass}
  \begin{fivenum}
    0 & 0.092 & 0.15 & 0.211 & 0.542 \\
  \end{fivenum}
\end{table}

\section{Conclusions}\label{sec:conclusions}

In this paper we have illustrated the results of the first
Italian~\acl{ASN}, that is required since~2010 to apply for a
permanent position as associate or full professor at Italian
universities. The scientific profile of applicants has been evaluated
also using three quantitative, paper-counting and citation-based
indicators. Qualification had to be preferably given to those
applicants whose quantitative indicators exceeded some pre-computed
thresholds; the thresholds were computed as the medians of the same
indicators computed for tenured professors.

We have analyzed the results of the~\ac{ASN} at the global level and
at the level of individual scientific disciplines. At the global level
the~\ac{ASN} received~59,149 applications,
42.8\% of which were successful; this
percentage is quite modest, especially considering that the~\ac{ASN}
does not grant professorship positions, but only allows qualified
individuals to apply to future openings. Although no direct comparison
with other countries is possible, it is instructive to observe that
the success rate in the 2013 French recruitment campaign for
professors and \emph{ma\^itres de conf\'erences} was 68.38\% (9,183
qualifications out of 13,430
applicants~\cite[p. 34]{Campagne13}). Looking at individual
disciplines, the fraction of successful qualifications ranges
from~15.4\%
to~81.1\%. In~25\% of
the~\acp{SD} the fraction of qualified applicants was less
than~35.3\%.
It would be tempting to attribute the poor performance of~\ac{ASN}
applicants to their low quality. Such an explanation would be
unsatisfactory, since~77.2\% of the applications come
from over-median applicants, but only~52.8\% of
them got qualification.

Over-median applicants were, on average, more likely to get
qualification than under-median ones. However, the distribution of the
conditional qualification probabilities for over-median applicants
spans a large range, with some disciplines granting qualification to a
small fraction of over-median applicants, and others qualifying most
of them. Half of the examination committees denied qualification to
more than half of the applicants that satisfied the quantitative
requirements for qualification. Therefore, exceeding the medians is
poorly correlated with getting qualification in half of the~\acp{SD},
suggesting that the role of the indicators should be reconsidered.

The analysis of the medians reveals some interesting facts. First,
there are disciplines where the thresholds for associate professor
qualification are higher than those for full professor
qualification. This means that in those disciplines it is easier to
pass the quantitative requirements for the higher academic rank, than
it is to pass the requirements for the lower rank. The second
observation is that there are medians equal to zero in several
non-bibliometric disciplines. This implies that the corresponding
indicators are zero for at least half of the tenured professors,
making those indicators not very useful for assessing the scientific
profile of applicants. Finally, medians for non-bibliometric
disciplines are not pairwise correlated, as it would be reasonable to
expect (and as it happens for bibliometric disciplines). These
observations suggest possible issues either in the computation of the
medians, or in the definition of inappropriate quantitative
indicators.

The data we have examined in this paper show \emph{what} happened, but
do not provide sufficient information to explain \emph{why} it
happened. More insights might come from the analysis of the curricula
of applicants and the final reports written by the examination
committees. We are planning to use natural language processing and
text analysis techniques to efficiently analyze the large body of
unstructured text documents produced by the~\ac{ASN}, and we will
report the outcomes of such analysis in a future work.

Another source of useful information is represented by the values of
the three quantitative indicators of applicants. The availability of
bibliometric information for a large population of researchers in all
scientific fields represents a valuable dataset for studying how
quantitative indicators behave across all scientific areas.

\appendix

\section{Notation}\label{all:notation}

\begin{tabular}{rcl}
  $M_1, M_2, M_3$  & := & Medians \\
  $B_1, B_2, B_3$  & := & Bibliometric indicators \\
  $N_1, N_2, N_3$  & := & Non-bibliometric indicators \\
  $\mathit{ind}_1, \mathit{ind}_2, \mathit{ind}_3$ & := & Quantitative indicators (either bibliometric or non-bibliometric) \\
  $\NA$         & := & Number of applications \\
  $\PQ$         & := & Fraction of qualified applicants \\
  $\PQU$        & := & Fraction of under-median applicants that got qualification \\
  $\PQO$        & := & Fraction of over-median applicants that got qualification \\
  $\PVR$        & := & Pareto Violation Ratio \\
\end{tabular}\medskip

The suffix $.F$ indicates that the variable refers to full professor
applicants, while the suffix $.A$ indicates that the variable refers
to associate professor applicants.

\section{List of Scientific Disciplines}\label{app:list-sd}

The list below enumerates all scientific areas (first indentation
level), macro-sectors (second indentation level) and scientific
disciplines, as they are defined when the~\ac{ASN} started~\cite[Annex
  A]{dm-set-concorsuali}.  We use here the English translation
provided by the Italian National University Council
(CUN)\footnote{\url{https://www.cun.it/documentazione/academic-fields-and-disciplines-list/},
  accessed on 2014-08-19}, since all names are officially defined in
Italian only.

\begin{multicols}{2}
\begin{scriptsize}
\begin{description}\item[01] Mathematics and computer sciences
\begin{description}
\item[01/A] Mathematics
\begin{description}
\item[01/A1] Mathematical logic, mathematics education and history of mathematics
\item[01/A2] Geometry and algebra
\item[01/A3] Mathematical analysis, probability and statistics
\item[01/A4] Mathematical physics
\item[01/A5] Numerical analysis
\item[01/A6] Operational research
\end{description}
\item[01/B] Informatics
\begin{description}
\item[01/B1] Informatics
\end{description}
\end{description}
\item[02] Physics
\begin{description}
\item[02/A] Physics of fundamental interactions
\begin{description}
\item[02/A1] Experimental physics of fundamental interactions
\item[02/A2] Theoretical physics of fundamental interactions
\end{description}
\item[02/B] Physics of matter
\begin{description}
\item[02/B1] Experimental physics of matter
\item[02/B2] Theoretical physics of matter
\item[02/B3] Applied physics
\end{description}
\item[02/C] Astronomy, astrophysics, Earth and planetary physics
\begin{description}
\item[02/C1] Astronomy, astrophysics, Earth and planetary physics
\end{description}
\end{description}
\item[03] Chemistry
\begin{description}
\item[03/A] Analytical and physical chemistry
\begin{description}
\item[03/A1] Analytical chemistry
\item[03/A2] Models and methods for chemistry
\end{description}
\item[03/B] Inorganic chemistry and applied technologies
\begin{description}
\item[03/B1] Principles of chemistry and inorganic systems
\item[03/B2] Chemical basis of technology applications
\end{description}
\item[03/C] Organic, industrial and applied chemistry
\begin{description}
\item[03/C1] Organic chemistry
\item[03/C2] Industrial and applied chemistry
\end{description}
\item[03/D] Medicinal and food chemistry and applied technologies
\begin{description}
\item[03/D1] Medicinal, toxicological and nutritional chemistry and applied technologies
\item[03/D2] Drug technology, socioeconomics and regulations
\end{description}
\end{description}
\item[04] Earth sciences
\begin{description}
\item[04/A] Earth sciences
\begin{description}
\item[04/A1] Geochemistry, mineralogy, petrology, volcanology, Earth resources and applications
\item[04/A2] Structural geology, stratigraphy, sedimentology and paleontology
\item[04/A3] Applied geology, physical geography and geomorphology
\item[04/A4] Geophysics
\end{description}
\end{description}
\item[05] Biology
\begin{description}
\item[05/A] Plant biology
\begin{description}
\item[05/A1] Botany
\item[05/A2] Plant physiology
\end{description}
\item[05/B] Animal biology and anthropology
\begin{description}
\item[05/B1] Zoology and anthropology
\item[05/B2] Comparative anatomy and cytology
\end{description}
\item[05/C] Ecology
\begin{description}
\item[05/C1] Ecology
\end{description}
\item[05/D] Physiology
\begin{description}
\item[05/D1] Physiology
\end{description}
\item[05/E] Experimental and clinical biochemistry and molecular biology
\begin{description}
\item[05/E1] General biochemistry and clinical biochemistry
\item[05/E2] Molecular biology
\end{description}
\item[05/F] Experimental biology
\begin{description}
\item[05/F1] Experimental biology
\end{description}
\item[05/G] Experimental and clinical pharmacology
\begin{description}
\item[05/G1] Pharmacology, clinical pharmacology and pharmacognosy
\end{description}
\item[05/H] Human anatomy and histology
\begin{description}
\item[05/H1] Human anatomy
\item[05/H2] Histology
\end{description}
\item[05/I] Genetics and microbiology
\begin{description}
\item[05/I1] Genetics and microbiology
\end{description}
\end{description}
\item[06] Medicine
\begin{description}
\item[06/A] Pathology and laboratory medicine
\begin{description}
\item[06/A1] Medical genetics
\item[06/A2] Experimental medicine, pathophysiology and clinical pathology
\item[06/A3] Microbiology and clinical microbiology
\item[06/A4] Pathology
\end{description}
\item[06/B] General clinical medicine
\begin{description}
\item[06/B1] Internal medicine
\end{description}
\item[06/C] General clinical surgery
\begin{description}
\item[06/C1] General surgery
\end{description}
\item[06/D] Specialized clinical medicine
\begin{description}
\item[06/D1] Cardiovascular and respiratory diseases
\item[06/D2] Endocrinology, nephrology, food and wellness sciences
\item[06/D3] Blood diseases, oncology and rheumatology
\item[06/D4] Skin, contagious and gastrointestinal diseases
\item[06/D5] Psychiatry
\item[06/D6] Neurology
\end{description}
\item[06/E] Specialized clinical surgery
\begin{description}
\item[06/E1] Heart, thoracic and vascular surgery
\item[06/E2] Plastic and paediatric surgery and urology
\item[06/E3] Neurosurgery and maxillofacial surgery
\end{description}
\item[06/F] Integrated clinical surgery
\begin{description}
\item[06/F1] Odontostomatologic diseases
\item[06/F2] Eye diseases
\item[06/F3] Otorhinolaryngology and audiology
\item[06/F4] Musculoskeletal diseases and physical and rehabilitation medicine
\end{description}
\item[06/G] Paediatrics
\begin{description}
\item[06/G1] Paediatrics and child neuropsychiatry
\end{description}
\item[06/H] Gynaecology
\begin{description}
\item[06/H1] Obstetrics and gynecology
\end{description}
\item[06/I] Radiology
\begin{description}
\item[06/I1] Diagnostic imaging, radiotherapy and neuroradiology
\end{description}
\item[06/L] Anaesthesiology
\begin{description}
\item[06/L1] Anaesthesiology
\end{description}
\item[06/M] Public health
\begin{description}
\item[06/M1] Hygiene, public health, nursing and medical statistics
\item[06/M2] Forensic and occupational medicine
\end{description}
\item[06/N] Applied medical technologies
\begin{description}
\item[06/N1] Applied medical technologies
\end{description}
\end{description}
\item[07] Agricultural and veterinary sciences
\begin{description}
\item[07/A] Agricultural economics and appraisal
\begin{description}
\item[07/A1] Agricultural economics and appraisal
\end{description}
\item[07/B] Agricultural and forest systems
\begin{description}
\item[07/B1] Agronomy and field, vegetable, ornamental cropping systems
\item[07/B2] Arboriculture and forest systems
\end{description}
\item[07/C] Agricultural, forest and biosytems engineering
\begin{description}
\item[07/C1] Agricultural, forest and biosystems engineering
\end{description}
\item[07/D] Plant pathology and entomology
\begin{description}
\item[07/D1] Plant pathology and entomology
\end{description}
\item[07/E] Agricultural chemistry and agricultural genetics
\begin{description}
\item[07/E1] Agricultural chemistry, agricultural genetics and pedology
\end{description}
\item[07/F] Food technology and agricultural microbiology
\begin{description}
\item[07/F1] Food science and technology
\item[07/F2] Agricultural microbiology
\end{description}
\item[07/G] Animal science and technology
\begin{description}
\item[07/G1] Animal science and technology
\end{description}
\item[07/H] Veterinary medicine
\begin{description}
\item[07/H1] Veterinary anatomy and physiology
\item[07/H2] Veterinary pathology and inspection of foods of animal origin
\item[07/H3] Infectious and parasitic animal diseases
\item[07/H4] Clinical veterinary medicine and pharmacology
\item[07/H5] Clinical veterinary surgery and obstetrics
\end{description}
\end{description}
\item[08] Civil engineering and architecture
\begin{description}
\item[08/A] Landscape and infrastructural engineering
\begin{description}
\item[08/A1] Hydraulics, hydrology, hydraulic and marine constructions
\item[08/A2] Sanitary and environmental engineering, hydrocarbons and underground fluids, safety and protection engineering
\item[08/A3] Infrastructural and transportation engineering, real estate appraisal and investment valuation
\item[08/A4] Geomatics
\end{description}
\item[08/B] Structural and geotechnical engineering
\begin{description}
\item[08/B1] Geotechnics
\item[08/B2] Structural mechanics
\item[08/B3] Structural engineering
\end{description}
\item[08/C] Design and technological planning of architecture
\begin{description}
\item[08/C1] Design and technological planning of architecture
\end{description}
\item[08/D] Architectural design
\begin{description}
\item[08/D1] Architectural design
\end{description}
\item[08/E] Drawing, architectural restoration and history
\begin{description}
\item[08/E1] Drawing
\item[08/E2] Architectural restoration and history
\end{description}
\item[08/F] Urban and landscape planning and design
\begin{description}
\item[08/F1] Urban and landscape planning and design
\end{description}
\end{description}
\item[09] Industrial and information engineering
\begin{description}
\item[09/A] Mechanical and aerospace engineering and naval architecture
\begin{description}
\item[09/A1] Aeronautical and aerospace engineering and naval architecture
\item[09/A2] Applied mechanics
\item[09/A3] Industrial design, machine construction and metallurgy
\end{description}
\item[09/B] Manufacturing, industrial and managenent engineering
\begin{description}
\item[09/B1] Manufacturing technology and systems
\item[09/B2] Industrial mechanical plants
\item[09/B3] Business and management engineering
\end{description}
\item[09/C] Energy, thermomechanical and nuclear engineering
\begin{description}
\item[09/C1] Fluid machinery, energy systems and power generation
\item[09/C2] Technical physics and nuclear engineering
\end{description}
\item[09/D] Chemical and materials engineering
\begin{description}
\item[09/D1] Materials science and technology
\item[09/D2] Systems, methods and technologies of chemical and process engineering
\item[09/D3] Chemical plants and technologies
\end{description}
\item[09/E] Electrical and electronic engineering and measurements
\begin{description}
\item[09/E1] Electrical technology
\item[09/E2] Electrical energy engineering
\item[09/E3] Electronics
\item[09/E4] Measurements
\end{description}
\item[09/F] Telecommunications engineering and electromagnetic fields
\begin{description}
\item[09/F1] Electromagnetic fields
\item[09/F2] Telecommunications
\end{description}
\item[09/G] Systems engineering and bioengineering
\begin{description}
\item[09/G1] Systems and control engineering
\item[09/G2] Bioengineering
\end{description}
\item[09/H] Computer engineering
\begin{description}
\item[09/H1] Information processing systems
\end{description}
\end{description}
\item[10] Antiquities, philology, literary studies, art history
\begin{description}
\item[10/A] Archaeological sciences
\begin{description}
\item[10/A1] Archaeology
\end{description}
\item[10/B] Art history
\begin{description}
\item[10/B1] Art history
\end{description}
\item[10/C] Cinema, music, performing arts, television and media studies
\begin{description}
\item[10/C1] Cinema, music, performing arts, television and media studies
\end{description}
\item[10/D] Sciences of antiquity
\begin{description}
\item[10/D1] Ancient history
\item[10/D2] Greek language and literature
\item[10/D3] Latin language and literature
\item[10/D4] Classical and late antique philology
\end{description}
\item[10/E] Medieval latin and romance philologies and literatures
\begin{description}
\item[10/E1] Medieval latin and romance philologies and literatures
\end{description}
\item[10/F] Italian studies and comparative literatures
\begin{description}
\item[10/F1] Italian literature, literary criticism and comparative literature
\item[10/F2] Contemporary Italian literature
\item[10/F3] Italian linguistics and philology
\end{description}
\item[10/G] Glottology and linguistics
\begin{description}
\item[10/G1] Glottology and linguistics
\end{description}
\item[10/H] French studies
\begin{description}
\item[10/H1] French language, literature and culture
\end{description}
\item[10/I] Spanish and Hispanic studies
\begin{description}
\item[10/I1] Spanish and Hispanic languages, literatures and cultures
\end{description}
\item[10/L] English and Anglo-American studies
\begin{description}
\item[10/L1] English and Anglo-American languages, literatures and cultures
\end{description}
\item[10/M] Germanic and Slavic languages, literatures and cultures
\begin{description}
\item[10/M1] Germanic languages, literatures and cultures
\item[10/M2] Slavic studies
\end{description}
\item[10/N] Eastern cultures
\begin{description}
\item[10/N1] Ancient Near Eastern, Middle Eastern and African cultures
\item[10/N3] Central and East Asian cultures
\end{description}
\end{description}
\item[11] History, philosophy, pedagogy and psychology
\begin{description}
\item[11/A] History
\begin{description}
\item[11/A1] Medieval history
\item[11/A2] Modern history
\item[11/A3] Contemporary history
\item[11/A4] Science of books and documents, history of religions
\item[11/A5] Demography, ethnography and anthropology
\end{description}
\item[11/B] Geography
\begin{description}
\item[11/B1] Geography
\end{description}
\item[11/C] Philosophy
\begin{description}
\item[11/C1] Theoretical philosophy
\item[11/C2] Logic, history and philosophy of science
\item[11/C3] Moral philosophy
\item[11/C4] Aesthetics and philosophy of languages
\item[11/C5] History of philosophy
\end{description}
\item[11/D] Educational theories
\begin{description}
\item[11/D1] Educational theories and history of educational theories
\item[11/D2] Methodologies of teaching, special education and educational research
\end{description}
\item[11/E] Psychology
\begin{description}
\item[11/E1] General psychology, psychobiology and psychometrics
\item[11/E2] Developmental and educational psychology
\item[11/E3] Social psychology and work and organizational psychology
\item[11/E4] Clinical and dynamic psychology
\end{description}
\end{description}
\item[12] Law studies
\begin{description}
\item[12/A] Private law
\begin{description}
\item[12/A1] Private law
\end{description}
\item[12/B] Business, navigation and air law and labour law
\begin{description}
\item[12/B1] Business, navigation and air law
\item[12/B2] Labour law
\end{description}
\item[12/C] Constitutional and ecclesiastical law
\begin{description}
\item[12/C1] Constitutional law
\item[12/C2] Ecclesiastical law and canon law
\end{description}
\item[12/D] Administrative and tax law
\begin{description}
\item[12/D1] Administrative law
\item[12/D2] Tax law
\end{description}
\item[12/E] International and European Union law, comparative, economics and markets law
\begin{description}
\item[12/E1] International and European Union law
\item[12/E2] Comparative law
\item[12/E3] Economics, financial and agri-food markets law and regulation
\end{description}
\item[12/F] Civil procedural law
\begin{description}
\item[12/F1] Civil procedural law
\end{description}
\item[12/G] Criminal law and criminal procedure
\begin{description}
\item[12/G1] Criminal law
\item[12/G2] Criminal procedure
\end{description}
\item[12/H] Roman law, history of medieval and modern law and philosophy of law
\begin{description}
\item[12/H1] Roman and ancient law
\item[12/H2] History of medieval and modern law
\item[12/H3] Philosophy of law
\end{description}
\end{description}
\item[13] Economics and statistics
\begin{description}
\item[13/A] Economics
\begin{description}
\item[13/A1] Economics
\item[13/A2] Economic policy
\item[13/A3] Public economics
\item[13/A4] Applied economics
\item[13/A5] Econometrics
\end{description}
\item[13/B] Business administration and Management
\begin{description}
\item[13/B1] Business administration and Management
\item[13/B2] Management
\item[13/B3] Organization studies
\item[13/B4] Financial Markets and Institutions
\item[13/B5] Commodity science
\end{description}
\item[13/C] Economic history
\begin{description}
\item[13/C1] Economic history
\end{description}
\item[13/D] Statistics and mathematical methods for decisions
\begin{description}
\item[13/D1] Statistics
\item[13/D2] Economic statistics
\item[13/D3] Demography and social statistics
\item[13/D4] Mathematical methods of economics, finance and actuarial sciences
\end{description}
\end{description}
\item[14] Political and social sciences
\begin{description}
\item[14/A] Political theory
\begin{description}
\item[14/A1] Political philosophy
\item[14/A2] Political science
\end{description}
\item[14/B] Political history
\begin{description}
\item[14/B1] History of political thought and institutions
\item[14/B2] History of international relations and of non-European societies and institutions
\end{description}
\item[14/C] Sociology
\begin{description}
\item[14/C1] General and political sociology, sociology of law
\item[14/C2] Sociology of culture and communication
\end{description}
\item[14/D] Applied sociology
\begin{description}
\item[14/D1] Sociology of economy and labour, sociology of land and environment\end{description}
\end{description}
\end{description}\end{scriptsize}
\end{multicols}

\section{Basic Statistics}\label{app:statistics}

Table~\ref{tab:basic-statistics} shows basic statistics for
each~\acp{SD}: number of applicants, number and fraction of
qualifications, fraction of qualified over- and under-median
applicants ($\PQO, PQU$). Note that $\PQU$ is undefined for a couple
of disciplines where there are no under-median applicants for the full
professor role. In these cases the reported value is ``NaN''.
  
\begin{scriptsize}
\begin{longtable}{@{\extracolsep{\fill}}lllllllllll}
  \caption{Basic Statistics of the~ASN\label{tab:basic-statistics}}\\
\toprule
{\em Sc. Discipline} & \multicolumn{5}{l}{\em Full Professor} & \multicolumn{5}{l}{\em Associate Professor} \\
\cmidrule{2-6} \cmidrule{7-11}
& App. & \em{Qualified} & $\PQ$ & $\PQO$ & $\PQU$ & App. & \em{Qualified} & $\PQ$ & $\PQO$ & $\PQU$ \\
\midrule
\endhead
\multicolumn{11}{l}{\em Area 1: Mathematics and computer sciences (MCS)}\\
01/A1 & 76 & 30 & 0.395 & 0.472 & 0.217 & 103 & 57 & 0.553 & 0.600 & 0.500\\
01/A2 & 126 & 51 & 0.405 & 0.610 & 0.023 & 222 & 123 & 0.554 & 0.745 & 0.164\\
01/A3 & 185 & 102 & 0.551 & 0.750 & 0.148 & 291 & 144 & 0.495 & 0.717 & 0.096\\
01/A4 & 144 & 62 & 0.431 & 0.455 & 0.304 & 231 & 91 & 0.394 & 0.446 & 0.232\\
01/A5 & 40 & 16 & 0.400 & 0.433 & 0.300 & 93 & 36 & 0.387 & 0.492 & 0.143\\
01/A6 & 34 & 15 & 0.441 & 0.517 & 0.000 & 49 & 23 & 0.469 & 0.769 & 0.130\\
01/B1 & 306 & 80 & 0.261 & 0.356 & 0.023 & 592 & 240 & 0.405 & 0.642 & 0.022\\
\multicolumn{11}{l}{\em Area 2: Physics (PHY)}\\
02/A1 & 356 & 212 & 0.596 & 0.651 & 0.375 & 549 & 412 & 0.750 & 0.842 & 0.362\\
02/A2 & 239 & 170 & 0.711 & 0.877 & 0.115 & 304 & 250 & 0.822 & 0.948 & 0.543\\
02/B1 & 230 & 160 & 0.696 & 0.796 & 0.327 & 506 & 376 & 0.743 & 0.904 & 0.321\\
02/B2 & 139 & 93 & 0.669 & 0.951 & 0.263 & 330 & 209 & 0.633 & 0.949 & 0.351\\
02/B3 & 280 & 70 & 0.250 & 0.276 & 0.145 & 696 & 194 & 0.279 & 0.330 & 0.170\\
02/C1 & 207 & 55 & 0.266 & 0.353 & 0.019 & 536 & 235 & 0.438 & 0.548 & 0.018\\
\multicolumn{11}{l}{\em Area 3: Chemistry (CHE)}\\
03/A1 & 57 & 29 & 0.509 & 0.703 & 0.150 & 224 & 118 & 0.527 & 0.713 & 0.149\\
03/A2 & 160 & 72 & 0.450 & 0.523 & 0.156 & 313 & 152 & 0.486 & 0.640 & 0.110\\
03/B1 & 143 & 119 & 0.832 & 0.991 & 0.148 & 266 & 185 & 0.695 & 1.000 & 0.120\\
03/B2 & 69 & 56 & 0.812 & 0.964 & 0.214 & 213 & 165 & 0.775 & 1.000 & 0.186\\
03/C1 & 104 & 46 & 0.442 & 0.548 & 0.000 & 205 & 97 & 0.473 & 0.681 & 0.071\\
03/C2 & 45 & 19 & 0.422 & 0.475 & 0.000 & 85 & 43 & 0.506 & 0.714 & 0.103\\
03/D1 & 93 & 32 & 0.344 & 0.449 & 0.042 & 264 & 127 & 0.481 & 0.646 & 0.140\\
03/D2 & 24 & 14 & 0.583 & 0.923 & 0.182 & 79 & 47 & 0.595 & 0.843 & 0.143\\
\multicolumn{11}{l}{\em Area 4: Earth sciences (EAS)}\\
04/A1 & 98 & 43 & 0.439 & 0.597 & 0.000 & 170 & 96 & 0.565 & 0.770 & 0.042\\
04/A2 & 96 & 43 & 0.448 & 0.489 & 0.000 & 195 & 124 & 0.636 & 0.780 & 0.156\\
04/A3 & 72 & 28 & 0.389 & 0.483 & 0.000 & 219 & 87 & 0.397 & 0.463 & 0.000\\
04/A4 & 134 & 34 & 0.254 & 0.282 & 0.059 & 247 & 59 & 0.239 & 0.337 & 0.048\\
\multicolumn{11}{l}{\em Area 5: Biology (BIO)}\\
05/A1 & 118 & 49 & 0.415 & 0.480 & 0.000 & 284 & 112 & 0.394 & 0.478 & 0.019\\
05/A2 & 35 & 17 & 0.486 & 0.548 & 0.000 & 64 & 30 & 0.469 & 0.667 & 0.000\\
05/B1 & 108 & 33 & 0.306 & 0.398 & 0.000 & 350 & 117 & 0.334 & 0.427 & 0.000\\
05/B2 & 87 & 30 & 0.345 & 0.435 & 0.000 & 210 & 69 & 0.329 & 0.425 & 0.020\\
05/C1 & 80 & 58 & 0.725 & 0.853 & 0.000 & 230 & 121 & 0.526 & 0.871 & 0.000\\
05/D1 & 151 & 56 & 0.371 & 0.455 & 0.146 & 382 & 130 & 0.340 & 0.411 & 0.140\\
05/E1 & 320 & 170 & 0.531 & 0.660 & 0.045 & 695 & 374 & 0.538 & 0.766 & 0.097\\
05/E2 & 170 & 78 & 0.459 & 0.761 & 0.134 & 659 & 267 & 0.405 & 0.694 & 0.092\\
05/F1 & 189 & 156 & 0.825 & 0.914 & 0.459 & 563 & 389 & 0.691 & 0.921 & 0.301\\
05/G1 & 139 & 38 & 0.273 & 0.396 & 0.000 & 405 & 109 & 0.269 & 0.405 & 0.000\\
05/H1 & 84 & 38 & 0.452 & 0.551 & 0.000 & 173 & 69 & 0.399 & 0.496 & 0.095\\
05/H2 & 64 & 16 & 0.250 & 0.302 & 0.000 & 145 & 28 & 0.193 & 0.292 & 0.000\\
05/I1 & 145 & 24 & 0.166 & 0.227 & 0.042 & 394 & 59 & 0.150 & 0.218 & 0.015\\
\multicolumn{11}{l}{\em Area 6: Medicine (MED)}\\
06/A1 & 57 & 29 & 0.509 & 0.722 & 0.143 & 179 & 86 & 0.480 & 0.813 & 0.136\\
06/A2 & 278 & 115 & 0.414 & 0.557 & 0.039 & 542 & 138 & 0.255 & 0.431 & 0.029\\
06/A3 & 90 & 18 & 0.200 & 0.281 & 0.000 & 193 & 47 & 0.244 & 0.367 & 0.000\\
06/A4 & 74 & 38 & 0.514 & 0.884 & 0.000 & 107 & 56 & 0.523 & 0.836 & 0.000\\
06/B1 & 241 & 73 & 0.303 & 0.514 & 0.000 & 449 & 140 & 0.312 & 0.496 & 0.000\\
06/C1 & 247 & 160 & 0.648 & 0.879 & 0.000 & 397 & 207 & 0.521 & 0.777 & 0.008\\
06/D1 & 125 & 33 & 0.264 & 0.418 & 0.000 & 311 & 123 & 0.395 & 0.641 & 0.000\\
06/D2 & 95 & 56 & 0.589 & 0.871 & 0.061 & 194 & 80 & 0.412 & 0.699 & 0.012\\
06/D3 & 122 & 37 & 0.303 & 0.600 & 0.016 & 306 & 127 & 0.415 & 0.632 & 0.044\\
06/D4 & 170 & 127 & 0.747 & 0.950 & 0.260 & 299 & 196 & 0.656 & 0.941 & 0.052\\
06/D5 & 45 & 22 & 0.489 & 0.611 & 0.000 & 106 & 54 & 0.509 & 0.783 & 0.000\\
06/D6 & 118 & 60 & 0.508 & 0.725 & 0.053 & 227 & 129 & 0.568 & 0.845 & 0.051\\
06/E1 & 96 & 28 & 0.292 & 0.426 & 0.057 & 193 & 72 & 0.373 & 0.508 & 0.119\\
06/E2 & 68 & 27 & 0.397 & 0.551 & 0.000 & 164 & 56 & 0.341 & 0.519 & 0.000\\
06/E3 & 59 & 23 & 0.390 & 0.444 & 0.214 & 116 & 67 & 0.578 & 0.815 & 0.029\\
06/F1 & 107 & 45 & 0.421 & 0.529 & 0.000 & 222 & 131 & 0.590 & 0.720 & 0.000\\
06/F2 & 41 & 10 & 0.244 & 0.385 & 0.000 & 81 & 14 & 0.173 & 0.269 & 0.000\\
06/F3 & 62 & 56 & 0.903 & 0.964 & 0.333 & 102 & 77 & 0.755 & 0.924 & 0.174\\
06/F4 & 138 & 41 & 0.297 & 0.373 & 0.000 & 218 & 62 & 0.284 & 0.354 & 0.000\\
06/G1 & 139 & 66 & 0.475 & 0.713 & 0.077 & 280 & 143 & 0.511 & 0.747 & 0.043\\
06/H1 & 84 & 40 & 0.476 & 0.909 & 0.000 & 174 & 101 & 0.580 & 0.856 & 0.000\\
06/I1 & 117 & 76 & 0.650 & 0.962 & 0.000 & 238 & 130 & 0.546 & 0.949 & 0.000\\
06/L1 & 64 & 11 & 0.172 & 0.306 & 0.000 & 125 & 39 & 0.312 & 0.487 & 0.000\\
06/M1 & 201 & 70 & 0.348 & 0.449 & 0.074 & 477 & 144 & 0.302 & 0.428 & 0.107\\
06/M2 & 95 & 31 & 0.326 & 0.470 & 0.000 & 190 & 76 & 0.400 & 0.526 & 0.105\\
06/N1 & 365 & 85 & 0.233 & 0.306 & 0.031 & 799 & 174 & 0.218 & 0.304 & 0.042\\
\multicolumn{11}{l}{\em Area 7: Agricultural sciences and veterinary medicine (AVM)}\\
07/A1 & 72 & 39 & 0.542 & 0.679 & 0.158 & 103 & 68 & 0.660 & 0.871 & 0.341\\
07/B1 & 51 & 33 & 0.647 & 0.767 & 0.000 & 125 & 41 & 0.328 & 0.390 & 0.000\\
07/B2 & 57 & 33 & 0.579 & 0.732 & 0.188 & 149 & 69 & 0.463 & 0.553 & 0.038\\
07/C1 & 47 & 24 & 0.511 & 0.571 & 0.000 & 86 & 31 & 0.360 & 0.484 & 0.000\\
07/D1 & 59 & 44 & 0.746 & 0.808 & 0.286 & 165 & 88 & 0.533 & 0.678 & 0.170\\
07/E1 & 76 & 36 & 0.474 & 0.654 & 0.083 & 170 & 102 & 0.600 & 0.702 & 0.326\\
07/F1 & 42 & 15 & 0.357 & 0.455 & 0.000 & 137 & 56 & 0.409 & 0.478 & 0.267\\
07/F2 & 25 & 14 & 0.560 & 0.778 & 0.000 & 60 & 39 & 0.650 & 0.791 & 0.294\\
07/G1 & 56 & 34 & 0.607 & 0.756 & 0.000 & 149 & 67 & 0.450 & 0.663 & 0.000\\
07/H1 & 41 & 27 & 0.659 & 1.000 & 0.000 & 59 & 31 & 0.525 & 0.939 & 0.000\\
07/H2 & 33 & 22 & 0.667 & 0.786 & 0.000 & 72 & 42 & 0.583 & 0.719 & 0.067\\
07/H3 & 33 & 21 & 0.636 & 0.840 & 0.000 & 77 & 53 & 0.688 & 0.897 & 0.053\\
07/H4 & 29 & 13 & 0.448 & 0.591 & 0.000 & 41 & 25 & 0.610 & 0.714 & 0.000\\
07/H5 & 29 & 18 & 0.621 & 0.643 & 0.000 & 50 & 35 & 0.700 & 0.854 & 0.000\\
\multicolumn{11}{l}{\em Area 8: Civil engineering and architecture (CEA)}\\
08/A1 & 63 & 22 & 0.349 & 0.367 & 0.000 & 138 & 63 & 0.457 & 0.521 & 0.000\\
08/A2 & 59 & 14 & 0.237 & 0.298 & 0.000 & 145 & 39 & 0.269 & 0.390 & 0.000\\
08/A3 & 69 & 24 & 0.348 & 0.385 & 0.235 & 117 & 40 & 0.342 & 0.493 & 0.125\\
08/A4 & 37 & 12 & 0.324 & 0.429 & 0.000 & 85 & 31 & 0.365 & 0.547 & 0.062\\
08/B1 & 34 & 20 & 0.588 & 0.692 & 0.250 & 60 & 38 & 0.633 & 0.762 & 0.333\\
08/B2 & 63 & 23 & 0.365 & 0.418 & 0.000 & 95 & 41 & 0.432 & 0.577 & 0.000\\
08/B3 & 60 & 25 & 0.417 & 0.481 & 0.000 & 95 & 36 & 0.379 & 0.493 & 0.000\\
08/C1 & 150 & 67 & 0.447 & 0.493 & 0.000 & 379 & 156 & 0.412 & 0.536 & 0.000\\
08/D1 & 180 & 37 & 0.206 & 0.240 & 0.000 & 548 & 112 & 0.204 & 0.252 & 0.000\\
08/E1 & 62 & 26 & 0.419 & 0.500 & 0.000 & 160 & 62 & 0.388 & 0.484 & 0.000\\
08/E2 & 134 & 44 & 0.328 & 0.355 & 0.077 & 394 & 97 & 0.246 & 0.293 & 0.029\\
08/F1 & 116 & 57 & 0.491 & 0.549 & 0.071 & 356 & 191 & 0.537 & 0.583 & 0.033\\
\multicolumn{11}{l}{\em Area 9: Industrial and information engineering (IIE)}\\
09/A1 & 87 & 48 & 0.552 & 0.672 & 0.217 & 163 & 77 & 0.472 & 0.679 & 0.020\\
09/A2 & 36 & 15 & 0.417 & 0.500 & 0.125 & 77 & 35 & 0.455 & 0.491 & 0.375\\
09/A3 & 80 & 45 & 0.562 & 0.589 & 0.286 & 159 & 103 & 0.648 & 0.787 & 0.353\\
09/B1 & 29 & 19 & 0.655 & 0.720 & 0.250 & 53 & 38 & 0.717 & 0.737 & 0.667\\
09/B2 & 13 & 7 & 0.538 & 0.636 & 0.000 & 51 & 41 & 0.804 & 0.862 & 0.727\\
09/B3 & 34 & 15 & 0.441 & 0.577 & 0.000 & 86 & 41 & 0.477 & 0.702 & 0.034\\
09/C1 & 64 & 44 & 0.688 & 0.679 & 0.750 & 125 & 59 & 0.472 & 0.525 & 0.378\\
09/C2 & 110 & 56 & 0.509 & 0.573 & 0.071 & 201 & 86 & 0.428 & 0.550 & 0.077\\
09/D1 & 181 & 33 & 0.182 & 0.224 & 0.000 & 463 & 95 & 0.205 & 0.266 & 0.000\\
09/D2 & 47 & 26 & 0.553 & 0.686 & 0.167 & 103 & 51 & 0.495 & 0.635 & 0.138\\
09/D3 & 49 & 29 & 0.592 & 0.725 & 0.000 & 67 & 37 & 0.552 & 0.756 & 0.136\\
09/E1 & 49 & 20 & 0.408 & 0.500 & 0.000 & 99 & 44 & 0.444 & 0.613 & 0.162\\
09/E2 & 66 & 48 & 0.727 & 0.828 & 0.000 & 83 & 46 & 0.554 & 0.737 & 0.154\\
09/E3 & 119 & 40 & 0.336 & 0.455 & 0.000 & 205 & 80 & 0.390 & 0.535 & 0.063\\
09/E4 & 61 & 24 & 0.393 & 0.436 & 0.000 & 94 & 30 & 0.319 & 0.391 & 0.167\\
09/F1 & 56 & 17 & 0.304 & 0.347 & 0.000 & 107 & 41 & 0.383 & 0.554 & 0.000\\
09/F2 & 98 & 36 & 0.367 & 0.424 & 0.000 & 153 & 56 & 0.366 & 0.539 & 0.020\\
09/G1 & 54 & 44 & 0.815 & 0.886 & 0.500 & 92 & 57 & 0.620 & 0.829 & 0.491\\
09/G2 & 80 & 29 & 0.362 & 0.453 & 0.000 & 168 & 63 & 0.375 & 0.562 & 0.063\\
09/H1 & 260 & 96 & 0.369 & 0.454 & 0.055 & 413 & 176 & 0.426 & 0.605 & 0.085\\
\multicolumn{11}{l}{\em Area 10: Antiquities, philology, literary studies, art history (APL)}\\
10/A1 & 160 & 90 & 0.562 & 0.621 & 0.000 & 553 & 324 & 0.586 & 0.669 & 0.000\\
10/B1 & 188 & 64 & 0.340 & 0.373 & 0.091 & 529 & 137 & 0.259 & 0.332 & 0.044\\
10/C1 & 142 & 62 & 0.437 & 0.513 & 0.111 & 429 & 150 & 0.350 & 0.446 & 0.057\\
10/D1 & 45 & 15 & 0.333 & 0.441 & 0.000 & 158 & 42 & 0.266 & 0.328 & 0.056\\
10/D2 & 67 & 46 & 0.687 & 0.714 & 0.250 & 185 & 121 & 0.654 & 0.708 & 0.000\\
10/D3 & 63 & 33 & 0.524 & 0.525 & 0.500 & 137 & 54 & 0.394 & 0.417 & 0.100\\
10/D4 & 92 & 52 & 0.565 & 0.584 & 0.000 & 206 & 88 & 0.427 & 0.466 & 0.000\\
10/E1 & 52 & 29 & 0.558 & 0.583 & 0.250 & 149 & 77 & 0.517 & 0.557 & 0.222\\
10/F1 & 186 & 54 & 0.290 & 0.319 & 0.050 & 476 & 115 & 0.242 & 0.285 & 0.025\\
10/F2 & 86 & 35 & 0.407 & 0.422 & 0.000 & 237 & 95 & 0.401 & 0.459 & 0.000\\
10/F3 & 79 & 41 & 0.519 & 0.577 & 0.000 & 194 & 136 & 0.701 & 0.743 & 0.000\\
10/G1 & 100 & 42 & 0.420 & 0.427 & 0.250 & 215 & 106 & 0.493 & 0.538 & 0.050\\
10/H1 & 42 & 29 & 0.690 & 0.690 & NaN & 158 & 104 & 0.658 & 0.757 & 0.045\\
10/I1 & 43 & 22 & 0.512 & 0.524 & 0.000 & 141 & 89 & 0.631 & 0.705 & 0.158\\
10/L1 & 127 & 79 & 0.622 & 0.676 & 0.250 & 305 & 179 & 0.587 & 0.637 & 0.156\\
10/M1 & 78 & 40 & 0.513 & 0.520 & 0.333 & 128 & 59 & 0.461 & 0.487 & 0.111\\
10/M2 & 44 & 15 & 0.341 & 0.389 & 0.125 & 73 & 29 & 0.397 & 0.460 & 0.000\\
10/N1 & 75 & 17 & 0.227 & 0.254 & 0.000 & 214 & 101 & 0.472 & 0.555 & 0.000\\
10/N3 & 49 & 31 & 0.633 & 0.646 & 0.000 & 119 & 76 & 0.639 & 0.680 & 0.421\\
\multicolumn{11}{l}{\em Area 11: History, philosophy, pedagogy and psychology (HPP)}\\
11/A1 & 49 & 19 & 0.388 & 0.452 & 0.000 & 172 & 60 & 0.349 & 0.420 & 0.000\\
11/A2 & 90 & 30 & 0.333 & 0.370 & 0.000 & 237 & 93 & 0.392 & 0.448 & 0.083\\
11/A3 & 116 & 40 & 0.345 & 0.374 & 0.000 & 425 & 173 & 0.407 & 0.441 & 0.000\\
11/A4 & 111 & 50 & 0.450 & 0.527 & 0.056 & 322 & 92 & 0.286 & 0.316 & 0.029\\
11/A5 & 55 & 23 & 0.418 & 0.449 & 0.167 & 158 & 59 & 0.373 & 0.457 & 0.000\\
11/B1 & 88 & 55 & 0.625 & 0.643 & 0.250 & 234 & 153 & 0.654 & 0.695 & 0.238\\
11/C1 & 63 & 29 & 0.460 & 0.529 & 0.167 & 217 & 90 & 0.415 & 0.457 & 0.000\\
11/C2 & 85 & 32 & 0.376 & 0.395 & 0.000 & 206 & 83 & 0.403 & 0.425 & 0.077\\
11/C3 & 89 & 32 & 0.360 & 0.378 & 0.267 & 286 & 118 & 0.413 & 0.460 & 0.059\\
11/C4 & 69 & 21 & 0.304 & 0.305 & 0.300 & 226 & 100 & 0.442 & 0.474 & 0.059\\
11/C5 & 122 & 49 & 0.402 & 0.432 & 0.091 & 399 & 144 & 0.361 & 0.398 & 0.000\\
11/D1 & 88 & 11 & 0.125 & 0.141 & 0.000 & 235 & 44 & 0.187 & 0.216 & 0.000\\
11/D2 & 92 & 15 & 0.163 & 0.205 & 0.000 & 237 & 46 & 0.194 & 0.238 & 0.000\\
11/E1 & 128 & 54 & 0.422 & 0.523 & 0.200 & 377 & 216 & 0.573 & 0.703 & 0.204\\
11/E2 & 65 & 17 & 0.262 & 0.362 & 0.000 & 190 & 46 & 0.242 & 0.316 & 0.070\\
11/E3 & 68 & 17 & 0.250 & 0.472 & 0.000 & 178 & 62 & 0.348 & 0.596 & 0.000\\
11/E4 & 113 & 15 & 0.133 & 0.209 & 0.022 & 319 & 53 & 0.166 & 0.217 & 0.043\\
\multicolumn{11}{l}{\em Area 12: Law (LAW)}\\
12/A1 & 136 & 45 & 0.331 & 0.354 & 0.000 & 312 & 84 & 0.269 & 0.311 & 0.022\\
12/B1 & 57 & 32 & 0.561 & 0.615 & 0.000 & 186 & 58 & 0.312 & 0.352 & 0.000\\
12/B2 & 34 & 13 & 0.382 & 0.433 & 0.000 & 78 & 25 & 0.321 & 0.352 & 0.000\\
12/C1 & 88 & 33 & 0.375 & 0.434 & 0.000 & 224 & 74 & 0.330 & 0.379 & 0.000\\
12/C2 & 20 & 9 & 0.450 & 0.529 & 0.000 & 48 & 22 & 0.458 & 0.525 & 0.125\\
12/D1 & 116 & 34 & 0.293 & 0.321 & 0.000 & 256 & 75 & 0.293 & 0.312 & 0.000\\
12/D2 & 31 & 11 & 0.355 & 0.440 & 0.000 & 84 & 16 & 0.190 & 0.229 & 0.000\\
12/E1 & 68 & 27 & 0.397 & 0.429 & 0.000 & 164 & 44 & 0.268 & 0.301 & 0.000\\
12/E2 & 97 & 25 & 0.258 & 0.272 & 0.000 & 186 & 78 & 0.419 & 0.453 & 0.000\\
12/E3 & 60 & 21 & 0.350 & 0.362 & 0.000 & 98 & 33 & 0.337 & 0.413 & 0.087\\
12/F1 & 29 & 8 & 0.276 & 0.320 & 0.000 & 65 & 18 & 0.277 & 0.316 & 0.000\\
12/G1 & 31 & 11 & 0.355 & 0.407 & 0.000 & 114 & 41 & 0.360 & 0.376 & 0.000\\
12/G2 & 32 & 13 & 0.406 & 0.433 & 0.000 & 87 & 42 & 0.483 & 0.539 & 0.091\\
12/H1 & 26 & 11 & 0.423 & 0.478 & 0.000 & 58 & 36 & 0.621 & 0.632 & 0.000\\
12/H2 & 26 & 11 & 0.423 & 0.450 & 0.333 & 55 & 32 & 0.582 & 0.612 & 0.333\\
12/H3 & 36 & 18 & 0.500 & 0.545 & 0.000 & 135 & 58 & 0.430 & 0.460 & 0.000\\
\multicolumn{11}{l}{\em Area 13: Economics and statistics (ECS)}\\
13/A1 & 310 & 136 & 0.439 & 0.458 & 0.000 & 463 & 208 & 0.449 & 0.479 & 0.146\\
13/A2 & 280 & 185 & 0.661 & 0.693 & 0.000 & 417 & 283 & 0.679 & 0.723 & 0.037\\
13/A3 & 85 & 49 & 0.576 & 0.583 & 0.000 & 140 & 68 & 0.486 & 0.508 & 0.125\\
13/A4 & 193 & 57 & 0.295 & 0.315 & 0.000 & 361 & 124 & 0.343 & 0.375 & 0.000\\
13/A5 & 41 & 23 & 0.561 & 0.531 & 0.667 & 69 & 33 & 0.478 & 0.479 & 0.476\\
13/B1 & 114 & 49 & 0.430 & 0.467 & 0.000 & 290 & 145 & 0.500 & 0.564 & 0.000\\
13/B2 & 141 & 47 & 0.333 & 0.348 & 0.000 & 288 & 112 & 0.389 & 0.444 & 0.000\\
13/B3 & 58 & 16 & 0.276 & 0.302 & 0.000 & 143 & 55 & 0.385 & 0.441 & 0.120\\
13/B4 & 99 & 59 & 0.596 & 0.608 & 0.000 & 169 & 103 & 0.609 & 0.669 & 0.000\\
13/B5 & 27 & 13 & 0.481 & 0.520 & 0.000 & 47 & 23 & 0.489 & 0.548 & 0.000\\
13/C1 & 81 & 40 & 0.494 & 0.533 & 0.000 & 145 & 78 & 0.538 & 0.609 & 0.000\\
13/D1 & 114 & 27 & 0.237 & 0.245 & 0.000 & 236 & 102 & 0.432 & 0.462 & 0.000\\
13/D2 & 63 & 27 & 0.429 & 0.429 & NaN & 113 & 41 & 0.363 & 0.390 & 0.000\\
13/D3 & 49 & 22 & 0.449 & 0.458 & 0.000 & 97 & 30 & 0.309 & 0.345 & 0.000\\
13/D4 & 100 & 37 & 0.370 & 0.381 & 0.000 & 120 & 46 & 0.383 & 0.407 & 0.000\\
\multicolumn{11}{l}{\em Area 14: Political and sociel sciences (PSS)}\\
14/A1 & 50 & 19 & 0.380 & 0.413 & 0.000 & 200 & 83 & 0.415 & 0.500 & 0.000\\
14/A2 & 65 & 14 & 0.215 & 0.237 & 0.000 & 153 & 55 & 0.359 & 0.404 & 0.000\\
14/B1 & 57 & 27 & 0.474 & 0.510 & 0.167 & 177 & 63 & 0.356 & 0.399 & 0.000\\
14/B2 & 42 & 21 & 0.500 & 0.526 & 0.250 & 156 & 99 & 0.635 & 0.664 & 0.200\\
14/C1 & 148 & 29 & 0.196 & 0.209 & 0.000 & 424 & 71 & 0.167 & 0.193 & 0.000\\
14/C2 & 84 & 31 & 0.369 & 0.413 & 0.000 & 283 & 55 & 0.194 & 0.247 & 0.000\\
14/D1 & 69 & 21 & 0.304 & 0.350 & 0.000 & 221 & 71 & 0.321 & 0.384 & 0.000\\\bottomrule
\end{longtable}
\end{scriptsize}

\section{Medians}\label{app:medians}

Table~\ref{tab:medians} contains the values of the medians for full
($M_1.F, M_2.F, M_3.F$) and associate professor qualification ($M_1.A,
M_2.A, M_3.A$). In some cases, \acp{SD} contain sub-disciplines with a
specific code (\emph{SSD}) and different medians.  Sub-disciplines may
have been defined for one role only, hence some values are
missing. ``B'' denotes bibliometric disciplines, ``NB'' denotes
non-bibliometric ones. The tag in the last column is to be interpreted
as follows: ``O'' or ``OO'' means that one or two medians for full
professor qualifications are zero; ``o'' or ``oo'' means that one or
two medians for associate professor qualification are zero; ``*''
means that medians for associate professor Pareto-dominate those for
full professor qualification.

\begin{scriptsize}
\begin{longtable}{@{\extracolsep{\fill}}llllllllll}
  \caption{Medians for full and associate professor qualification\label{tab:medians}}\\
\toprule
{\em Sc. Discipline} & {\em SSD} & {\em Bibliometric?} & \multicolumn{3}{l}{\em Full professor} & \multicolumn{3}{l}{\em Associate professor} & \\
\cmidrule{4-6}\cmidrule{7-9}
& & & $M_1.F$ & $M_1.F$ & $M_3.F$ & $M_1.A$ & $M_2.A$ & $M_3.A$ & \\
\midrule
\endhead
 01/A1 &  & B & 4.00 & 1.37 & 2.00 & 5.00 & 1.74 & 2.00 & * \\ 
  01/A1 & MAT/04 & B & 2.50 & 0.71 & 2.00 &  &  &  &  \\ 
  01/A2 &  & B & 9.00 & 3.23 & 3.00 & 8.00 & 1.65 & 2.00 &  \\ 
  01/A3 &  & B & 14.00 & 8.00 & 5.00 & 10.00 & 4.34 & 4.00 &  \\ 
  01/A4 &  & B & 15.00 & 8.78 & 5.00 & 13.00 & 5.95 & 4.00 &  \\ 
  01/A5 &  & B & 15.50 & 15.72 & 7.00 & 14.00 & 6.06 & 4.00 &  \\ 
  01/A6 &  & B & 17.00 & 12.83 & 6.50 & 17.00 & 9.35 & 6.00 &  \\ 
  01/B1 &  & B & 12.00 & 14.80 & 6.00 & 10.00 & 9.15 & 5.00 &  \\ 
  02/A1 &  & B & 78.00 & 105.03 & 22.00 & 59.50 & 104.08 & 18.00 &  \\ 
  02/A1 & FIS/01 & B & 55.00 & 67.06 & 14.00 & 44.00 & 42.59 & 11.00 &  \\ 
  02/A2 &  & B & 24.50 & 47.41 & 11.00 & 23.00 & 34.09 & 10.00 &  \\ 
  02/B1 &  & B & 54.00 & 46.72 & 12.00 & 38.00 & 32.09 & 9.00 &  \\ 
  02/B2 &  & B & 47.50 & 75.94 & 14.00 & 37.50 & 40.08 & 11.00 &  \\ 
  02/B2 & FIS/02 & B &  &  &  & 37.50 & 40.08 & 8.00 &  \\ 
  02/B2 & FIS/08 & B & 12.00 & 3.02 & 3.00 & 4.50 & 0.33 & 1.50 &  \\ 
  02/B3 &  & B & 43.50 & 34.63 & 10.00 & 27.00 & 22.47 & 8.00 &  \\ 
  02/C1 &  & B & 49.00 & 86.08 & 18.00 & 32.00 & 35.83 & 10.00 &  \\ 
  02/C1 & FIS/06 & B & 27.50 & 22.01 & 9.00 & 17.00 & 14.00 & 6.00 &  \\ 
  03/A1 &  & B & 41.00 & 53.81 & 12.00 & 26.00 & 29.47 & 9.00 &  \\ 
  03/A1 & CHIM/12 & B &  &  &  & 26.00 & 18.29 & 9.00 &  \\ 
  03/A2 &  & B & 42.50 & 46.01 & 11.00 & 34.50 & 34.27 & 10.00 &  \\ 
  03/B1 &  & B & 49.50 & 62.38 & 13.00 & 31.00 & 47.05 & 11.00 &  \\ 
  03/B2 &  & B & 42.00 & 59.89 & 12.00 & 24.50 & 31.70 & 9.00 &  \\ 
  03/C1 &  & B & 41.50 & 55.45 & 12.00 & 33.00 & 42.47 & 10.00 &  \\ 
  03/C2 &  & B & 37.00 & 50.01 & 12.00 & 35.00 & 38.77 & 11.00 &  \\ 
  03/D1 &  & B & 44.00 & 42.85 & 12.00 & 28.00 & 28.00 & 9.00 &  \\ 
  03/D2 &  & B & 43.00 & 48.60 & 12.00 & 24.00 & 35.30 & 10.00 &  \\ 
  04/A1 &  & B & 25.00 & 29.94 & 10.00 & 19.00 & 14.29 & 7.00 &  \\ 
  04/A1 & GEO/06 & B & 25.00 & 29.94 & 9.00 &  &  &  &  \\ 
  04/A1 & GEO/09 & B & 25.00 & 7.26 & 7.00 & 19.00 & 8.53 & 6.00 &  \\ 
  04/A2 &  & B & 17.00 & 15.31 & 8.00 & 13.00 & 8.74 & 6.00 &  \\ 
  04/A2 & GEO/02 & B & 17.00 & 15.31 & 7.00 &  &  &  &  \\ 
  04/A3 &  & B & 9.00 & 3.06 & 4.00 & 6.00 & 2.00 & 3.00 &  \\ 
  04/A4 &  & B & 19.00 & 14.18 & 6.00 & 17.00 & 13.66 & 6.00 &  \\ 
  04/A4 & GEO/11 & B & 11.50 & 6.61 & 5.00 & 17.00 & 6.33 & 4.50 &  \\ 
  05/A1 &  & B & 20.00 & 13.25 & 7.00 & 12.00 & 8.74 & 6.00 &  \\ 
  05/A1 & BIO/02 & B & 6.50 & 13.25 & 4.00 & 12.00 & 2.43 & 3.00 &  \\ 
  05/A2 &  & B & 20.00 & 35.47 & 10.00 & 14.00 & 24.45 & 8.50 &  \\ 
  05/B1 &  & B & 28.50 & 18.64 & 8.00 & 16.00 & 8.20 & 5.50 &  \\ 
  05/B1 & BIO/08 & B & 18.00 & 18.64 & 5.50 &  &  &  &  \\ 
  05/B2 &  & B & 26.00 & 22.96 & 9.00 & 18.50 & 16.66 & 7.00 &  \\ 
  05/C1 &  & B & 26.00 & 18.63 & 9.00 & 21.50 & 15.77 & 8.00 &  \\ 
  05/D1 &  & B & 25.00 & 44.88 & 12.00 & 19.00 & 26.47 & 9.00 &  \\ 
  05/E1 &  & B & 32.00 & 42.61 & 11.00 & 21.00 & 32.04 & 10.00 &  \\ 
  05/E2 &  & B & 37.00 & 80.32 & 15.00 & 17.00 & 40.64 & 11.00 &  \\ 
  05/F1 &  & B & 29.00 & 40.65 & 12.00 & 22.00 & 33.41 & 10.00 &  \\ 
  05/G1 &  & B & 41.50 & 75.58 & 15.00 & 24.50 & 38.34 & 11.00 &  \\ 
  05/G1 & BIO/15 & B & 30.00 & 32.06 & 10.50 & 24.50 & 15.94 & 8.00 &  \\ 
  05/H1 &  & B & 31.50 & 35.54 & 10.00 & 23.00 & 26.61 & 8.00 &  \\ 
  05/H2 &  & B & 28.00 & 42.35 & 12.00 & 23.00 & 38.26 & 11.00 &  \\ 
  05/I1 &  & B & 28.00 & 47.16 & 13.00 & 19.00 & 28.41 & 9.00 &  \\ 
  06/A1 &  & B & 62.50 & 113.33 & 17.50 & 30.50 & 66.18 & 13.00 &  \\ 
  06/A2 &  & B & 32.50 & 67.32 & 14.00 & 25.00 & 53.07 & 12.00 &  \\ 
  06/A2 & MED/02 & B & 8.00 & 1.00 & 3.00 & 25.00 & 1.66 & 2.50 &  \\ 
  06/A3 &  & B & 31.00 & 38.40 & 12.00 & 25.00 & 29.75 & 9.00 &  \\ 
  06/A4 &  & B & 59.00 & 88.18 & 15.00 & 41.00 & 37.60 & 12.00 &  \\ 
  06/B1 &  & B & 63.00 & 86.83 & 17.00 & 33.50 & 49.62 & 12.00 &  \\ 
  06/C1 &  & B & 25.00 & 13.36 & 7.00 & 15.00 & 8.33 & 6.00 &  \\ 
  06/D1 &  & B & 62.50 & 87.69 & 16.00 & 35.00 & 41.99 & 11.00 &  \\ 
  06/D2 &  & B & 69.00 & 106.36 & 18.00 & 39.00 & 52.19 & 12.00 &  \\ 
  06/D2 & MED/14 & B & 69.00 & 44.95 & 18.00 &  &  &  &  \\ 
  06/D2 & MED/49 & B & 29.50 & 80.12 & 13.00 &  &  &  &  \\ 
  06/D3 &  & B & 89.00 & 155.14 & 22.00 & 44.50 & 63.04 & 14.00 &  \\ 
  06/D3 & MED/16 & B &  &  &  & 44.50 & 43.24 & 12.50 &  \\ 
  06/D4 &  & B & 67.00 & 66.03 & 15.00 & 37.00 & 35.19 & 11.00 &  \\ 
  06/D4 & MED/35 & B & 67.00 & 38.26 & 10.50 &  &  &  &  \\ 
  06/D5 &  & B & 45.00 & 37.75 & 12.50 & 27.50 & 29.88 & 8.50 &  \\ 
  06/D6 &  & B & 78.00 & 82.76 & 16.00 & 43.00 & 60.70 & 13.00 &  \\ 
  06/E1 &  & B & 36.50 & 21.51 & 9.00 & 22.00 & 10.88 & 7.00 &  \\ 
  06/E2 &  & B & 42.83 & 16.85 & 7.50 & 26.00 & 12.15 & 6.00 &  \\ 
  06/E2 & MED/19 & B & 42.83 & 6.85 & 6.50 & 26.00 & 7.33 & 6.00 &  \\ 
  06/E3 &  & B & 36.00 & 22.72 & 8.00 & 21.50 & 11.25 & 6.00 &  \\ 
  06/E3 & MED/29 & B & 36.00 & 8.55 & 6.00 &  &  &  &  \\ 
  06/F1 &  & B & 21.00 & 8.16 & 6.00 & 10.00 & 2.78 & 4.00 &  \\ 
  06/F2 &  & B & 28.00 & 20.42 & 9.00 & 16.50 & 11.51 & 6.50 &  \\ 
  06/F3 &  & B & 28.00 & 13.96 & 7.00 & 16.00 & 7.36 & 5.00 &  \\ 
  06/F3 & MED/32 & B & 28.00 & 2.60 & 4.00 &  &  &  &  \\ 
  06/F4 &  & B & 17.75 & 9.42 & 6.00 & 14.50 & 3.55 & 4.00 &  \\ 
  06/G1 &  & B & 62.00 & 52.84 & 14.00 & 28.00 & 31.03 & 10.50 &  \\ 
  06/H1 &  & B & 57.00 & 50.80 & 14.00 & 23.00 & 20.75 & 8.00 &  \\ 
  06/I1 &  & B & 44.50 & 33.07 & 10.00 & 27.00 & 24.33 & 9.00 &  \\ 
  06/L1 &  & B & 31.00 & 25.57 & 10.00 & 18.00 & 14.06 & 7.00 &  \\ 
  06/M1 &  & B & 26.00 & 19.95 & 8.50 & 20.50 & 13.10 & 7.00 &  \\ 
  06/M1 & MED/45 & B &  &  &  & 20.50 & 13.10 & 3.00 &  \\ 
  06/M2 &  & B & 16.00 & 6.84 & 6.00 & 8.37 & 4.62 & 4.00 &  \\ 
  06/M2 & MED/43 & B & 16.00 & 6.84 & 4.00 &  &  &  &  \\ 
  06/N1 &  & B & 29.00 & 35.25 & 11.00 & 21.50 & 19.84 & 8.00 &  \\ 
  07/A1 &  & B & 2.00 & 0.05 & 1.00 & 2.50 & 0.34 & 1.00 & * \\ 
  07/B1 &  & B & 13.00 & 5.68 & 4.00 & 12.00 & 3.38 & 3.50 &  \\ 
  07/B2 &  & B & 15.00 & 9.64 & 6.00 & 8.00 & 3.48 & 4.00 &  \\ 
  07/B2 & AGR/06 & B &  &  &  & 4.50 & 0.69 & 2.00 &  \\ 
  07/C1 &  & B & 6.00 & 1.61 & 3.00 & 7.00 & 2.91 & 3.00 & * \\ 
  07/C1 & AGR/09 & B &  &  &  & 7.00 & 2.91 & 2.50 &  \\ 
  07/C1 & AGR/10 & B & 6.00 & 1.61 & 2.00 & 7.00 & 2.91 & 2.00 & * \\ 
  07/D1 &  & B & 15.00 & 5.75 & 6.00 & 12.00 & 4.40 & 5.00 &  \\ 
  07/D1 & AGR/11 & B &  &  &  & 12.00 & 4.40 & 3.00 &  \\ 
  07/E1 &  & B & 26.50 & 20.51 & 8.00 & 17.00 & 14.09 & 7.00 &  \\ 
  07/E1 & AGR/14 & B & 26.50 & 7.66 & 5.00 & 17.00 & 4.39 & 4.00 &  \\ 
  07/F1 &  & B & 25.00 & 19.13 & 8.00 & 20.00 & 15.89 & 8.00 &  \\ 
  07/F2 &  & B & 31.00 & 33.00 & 11.00 & 24.00 & 21.63 & 9.00 &  \\ 
  07/G1 &  & B & 23.00 & 10.36 & 6.00 & 17.00 & 5.73 & 5.00 &  \\ 
  07/G1 & AGR/18 & B &  &  &  & 17.00 & 5.73 & 4.00 &  \\ 
  07/H1 &  & B & 26.50 & 14.38 & 7.00 & 19.50 & 9.18 & 5.00 &  \\ 
  07/H2 &  & B & 22.00 & 8.77 & 5.50 & 20.00 & 7.64 & 5.00 &  \\ 
  07/H3 &  & B & 34.50 & 18.82 & 8.00 & 22.50 & 13.58 & 7.00 &  \\ 
  07/H4 &  & B & 16.00 & 7.28 & 5.00 & 17.00 & 6.47 & 5.00 &  \\ 
  07/H5 &  & B & 11.00 & 2.10 & 3.00 & 10.00 & 3.52 & 4.00 &  \\ 
  07/H5 & VET/09 & B & 11.00 & 2.10 & 2.00 &  &  &  &  \\ 
  08/A1 &  & B & 9.00 & 4.25 & 4.00 & 7.50 & 3.53 & 3.50 &  \\ 
  08/A2 &  & B & 9.00 & 5.10 & 4.00 & 10.00 & 3.02 & 4.00 &  \\ 
  08/A2 & ING-IND/28 & B & 3.00 & 0.15 & 2.00 &  &  &  &  \\ 
  08/A2 & ING-IND/29 & B &  &  &  & 6.00 & 3.02 & 4.00 &  \\ 
  08/A3 &  & B & 3.00 & 0.36 & 1.50 & 4.00 & 1.04 & 2.00 & * \\ 
  08/A3 & ICAR/04 & B &  &  &  & 4.00 & 1.04 & 1.00 &  \\ 
  08/A4 &  & B & 5.00 & 1.30 & 2.00 & 3.00 & 0.82 & 2.00 &  \\ 
  08/B1 &  & B & 5.00 & 4.09 & 4.00 & 5.00 & 2.54 & 3.00 &  \\ 
  08/B2 &  & B & 12.50 & 8.96 & 5.50 & 11.00 & 6.14 & 5.00 &  \\ 
  08/B3 &  & B & 9.00 & 4.40 & 4.00 & 8.00 & 3.18 & 4.00 &  \\ 
  08/C1 &  & NB & 2.00 & 19.00 & 0.00 & 2.00 & 16.00 & 0.00 & Oo \\ 
  08/D1 &  & NB & 2.00 & 19.00 & 0.00 & 2.00 & 12.00 & 0.00 & Oo \\ 
  08/D1 & ICAR/15 & NB &  &  &  & 0.50 & 12.00 & 0.00 & o \\ 
  08/E1 &  & NB & 1.00 & 25.00 & 0.00 & 2.00 & 15.00 & 0.00 & Oo \\ 
  08/E2 &  & NB & 1.50 & 21.00 & 0.00 & 1.00 & 16.00 & 0.00 & Oo \\ 
  08/F1 &  & NB & 2.00 & 23.00 & 0.00 & 1.00 & 14.00 & 0.00 & Oo \\ 
  08/F1 & ICAR/20 & NB & 1.00 & 23.00 & 0.00 &  &  &  & O \\ 
  09/A1 &  & B & 10.00 & 6.25 & 5.00 & 10.00 & 3.09 & 3.00 &  \\ 
  09/A1 & ING-IND/02 & B & 1.50 & 0.39 & 1.00 & 3.37 & 0.73 & 1.50 & * \\ 
  09/A1 & ING-IND/03 & B &  &  &  & 10.00 & 1.39 & 2.00 &  \\ 
  09/A2 &  & B & 9.00 & 5.47 & 5.00 & 7.00 & 5.72 & 4.00 &  \\ 
  09/A3 &  & B & 12.00 & 5.58 & 4.00 & 13.50 & 6.29 & 4.50 & * \\ 
  09/A3 & ING-IND/15 & B & 7.00 & 1.78 & 3.00 & 8.00 & 2.10 & 3.00 & * \\ 
  09/B1 &  & B & 14.00 & 7.31 & 5.00 & 13.00 & 7.38 & 5.00 &  \\ 
  09/B2 &  & B & 9.00 & 3.78 & 5.00 & 9.00 & 4.44 & 4.00 &  \\ 
  09/B3 &  & B & 7.00 & 7.00 & 5.00 & 8.00 & 4.60 & 4.00 &  \\ 
  09/C1 &  & B & 6.00 & 2.21 & 3.00 & 8.00 & 3.70 & 4.00 & * \\ 
  09/C2 &  & B & 8.00 & 3.82 & 4.00 & 11.00 & 3.09 & 4.00 &  \\ 
  09/C2 & ING-IND/11 & B & 4.00 & 1.62 & 3.00 &  &  &  &  \\ 
  09/D1 &  & B & 29.00 & 27.12 & 9.00 & 20.50 & 19.13 & 7.50 &  \\ 
  09/D1 & ING-IND/21 & B &  &  &  & 9.50 & 7.79 & 5.50 &  \\ 
  09/D2 &  & B & 34.00 & 27.91 & 9.00 & 18.00 & 18.53 & 8.00 &  \\ 
  09/D2 & ING-IND/26 & B &  &  &  & 18.00 & 18.53 & 7.00 &  \\ 
  09/D3 &  & B & 30.50 & 30.25 & 10.00 & 20.00 & 26.00 & 9.00 &  \\ 
  09/E1 &  & B & 18.00 & 13.95 & 6.00 & 18.00 & 11.88 & 6.00 &  \\ 
  09/E2 &  & B & 8.00 & 10.00 & 6.00 & 9.00 & 8.36 & 5.00 &  \\ 
  09/E2 & ING-IND/33 & B & 8.00 & 10.00 & 5.00 &  &  &  &  \\ 
  09/E3 &  & B & 26.00 & 23.58 & 8.00 & 23.00 & 17.11 & 7.00 &  \\ 
  09/E4 &  & B & 13.00 & 12.55 & 6.00 & 18.00 & 14.15 & 6.00 & * \\ 
  09/E4 & ING-IND/12 & B & 13.00 & 4.91 & 4.00 & 13.00 & 4.09 & 4.00 &  \\ 
  09/F1 &  & B & 33.00 & 21.14 & 8.00 & 23.00 & 17.14 & 8.00 &  \\ 
  09/F2 &  & B & 21.00 & 20.56 & 7.00 & 21.00 & 19.78 & 7.00 &  \\ 
  09/G1 &  & B & 18.00 & 21.02 & 8.00 & 18.00 & 20.00 & 8.00 &  \\ 
  09/G2 &  & B & 36.00 & 42.58 & 12.00 & 30.00 & 31.39 & 8.00 &  \\ 
  09/H1 &  & B & 13.00 & 13.48 & 6.00 & 10.00 & 10.67 & 6.00 &  \\ 
  10/A1 &  & NB & 1.00 & 23.00 & 3.00 & 1.00 & 17.00 & 1.00 &  \\ 
  10/A1 & L-ANT/01 & NB &  &  &  & 0.00 & 17.00 & 1.00 & o \\ 
  10/A1 & L-ANT/09 & NB & 0.50 & 23.00 & 3.00 &  &  &  &  \\ 
  10/A1 & L-FIL-LET/01 & NB &  &  &  & 0.00 & 17.00 & 1.00 & o \\ 
  10/B1 &  & NB & 1.00 & 24.00 & 0.00 & 2.00 & 17.00 & 0.00 & Oo \\ 
  10/C1 &  & NB & 2.00 & 19.00 & 1.00 & 2.00 & 16.00 & 1.00 &  \\ 
  10/C1 & L-ART/07 & NB &  &  &  & 2.00 & 16.00 & 0.00 & o \\ 
  10/C1 & L-ART/08 & NB &  &  &  & 2.00 & 13.00 & 0.00 & o \\ 
  10/D1 &  & NB & 1.00 & 19.00 & 3.00 & 1.00 & 14.00 & 2.00 &  \\ 
  10/D2 &  & NB & 1.00 & 18.00 & 3.00 & 1.00 & 12.00 & 1.00 &  \\ 
  10/D2 & L-FIL-LET/07 & NB &  &  &  & 0.00 & 9.00 & 1.00 & o \\ 
  10/D2 & L-LIN/20 & NB &  &  &  & 1.00 & 12.00 & 0.00 & o \\ 
  10/D3 &  & NB & 1.00 & 14.00 & 3.00 & 1.00 & 7.00 & 1.00 &  \\ 
  10/D4 &  & NB & 1.00 & 15.50 & 2.00 & 1.00 & 10.00 & 1.00 &  \\ 
  10/D4 & L-ANT/05 & NB &  &  &  & 0.00 & 10.00 & 1.00 & o \\ 
  10/D4 & L-FIL-LET/06 & NB & 1.00 & 15.50 & 0.00 & 1.00 & 10.00 & 0.00 & Oo \\ 
  10/E1 &  & NB & 2.00 & 14.00 & 3.00 & 1.00 & 14.00 & 2.00 &  \\ 
  10/F1 &  & NB & 2.00 & 18.00 & 2.00 & 2.00 & 15.00 & 2.00 &  \\ 
  10/F2 &  & NB & 2.00 & 16.00 & 2.00 & 2.00 & 16.50 & 1.00 &  \\ 
  10/F3 &  & NB & 2.00 & 19.50 & 2.00 & 2.00 & 14.50 & 1.00 &  \\ 
  10/F3 & L-FIL-LET/13 & NB & 1.00 & 19.50 & 2.00 & 1.50 & 14.50 & 1.00 &  \\ 
  10/G1 &  & NB & 1.00 & 20.00 & 0.00 & 1.00 & 18.00 & 0.00 & Oo \\ 
  10/H1 &  & NB & 1.00 & 15.00 & 1.00 & 2.00 & 10.50 & 0.00 & o \\ 
  10/H1 & L-LIN/04 & NB & 1.00 & 15.00 & 0.00 &  &  &  & O \\ 
  10/I1 &  & NB & 2.00 & 19.50 & 1.00 & 2.00 & 14.00 & 0.00 & o \\ 
  10/I1 & L-LIN/07 & NB &  &  &  & 1.00 & 14.00 & 0.00 & o \\ 
  10/L1 &  & NB & 1.00 & 15.00 & 1.00 & 1.00 & 12.00 & 1.00 &  \\ 
  10/M1 &  & NB & 1.00 & 15.00 & 1.00 & 1.00 & 12.00 & 0.00 & o \\ 
  10/M1 & L-FIL-LET/15 & NB & 0.00 & 15.00 & 0.00 &  &  &  & OO \\ 
  10/M2 &  & NB & 2.00 & 15.50 & 2.00 & 1.00 & 18.00 & 1.00 &  \\ 
  10/N1 &  & NB & 2.00 & 15.00 & 2.00 & 1.50 & 16.50 & 2.00 &  \\ 
  10/N1 & L-OR/01 & NB & 1.50 & 15.00 & 2.00 &  &  &  &  \\ 
  10/N1 & L-OR/04 & NB & 1.00 & 15.00 & 2.00 &  &  &  &  \\ 
  10/N1 & L-OR/10 & NB & 2.00 & 11.50 & 1.50 &  &  &  &  \\ 
  10/N1 & L-OR/12 & NB & 2.00 & 15.00 & 0.00 & 1.50 & 7.00 & 0.00 & Oo \\ 
  10/N3 &  & NB & 1.00 & 10.00 & 1.00 & 2.00 & 10.00 & 0.00 & o \\ 
  10/N3 & L-OR/21 & NB & 1.00 & 10.00 & 0.50 &  &  &  &  \\ 
  10/N3 & L-OR/22 & NB & 0.00 & 9.00 & 0.00 &  &  &  & OO \\ 
  11/A1 &  & NB & 2.00 & 19.00 & 1.00 & 2.00 & 17.50 & 0.00 & o \\ 
  11/A2 &  & NB & 2.00 & 18.00 & 0.00 & 2.00 & 12.00 & 0.00 & Oo \\ 
  11/A3 &  & NB & 2.00 & 16.00 & 0.00 & 1.71 & 11.50 & 0.00 & Oo \\ 
  11/A4 &  & NB & 2.00 & 18.00 & 0.00 & 1.00 & 12.00 & 0.00 & Oo \\ 
  11/A4 & M-STO/09 & NB & 1.00 & 18.00 & 0.00 &  &  &  & O \\ 
  11/A5 &  & NB & 2.00 & 17.00 & 1.00 & 2.00 & 12.50 & 1.00 &  \\ 
  11/B1 &  & NB & 1.00 & 14.00 & 0.00 & 1.00 & 12.00 & 0.00 & Oo \\ 
  11/C1 &  & NB & 4.00 & 21.50 & 1.00 & 2.50 & 13.50 & 0.00 & o \\ 
  11/C2 &  & NB & 1.11 & 18.00 & 1.00 & 2.00 & 10.00 & 1.00 &  \\ 
  11/C3 &  & NB & 3.00 & 21.50 & 1.00 & 2.00 & 14.00 & 0.00 & o \\ 
  11/C4 &  & NB & 3.00 & 23.00 & 1.00 & 2.00 & 19.00 & 0.50 &  \\ 
  11/C4 & M-FIL/05 & NB & 2.00 & 23.00 & 1.00 &  &  &  &  \\ 
  11/C5 &  & NB & 2.00 & 21.00 & 1.00 & 2.00 & 13.50 & 0.00 & o \\ 
  11/D1 &  & NB & 3.00 & 22.00 & 1.00 & 3.00 & 13.00 & 0.00 & o \\ 
  11/D2 &  & NB & 4.00 & 22.50 & 2.00 & 3.00 & 13.00 & 0.00 & o \\ 
  11/E1 &  & B & 23.00 & 19.33 & 9.00 & 14.00 & 11.84 & 6.00 &  \\ 
  11/E1 & M-PSI/03 & B & 23.00 & 7.89 & 9.00 & 4.50 & 11.84 & 3.00 &  \\ 
  11/E2 &  & B & 7.00 & 2.47 & 3.00 & 4.00 & 1.35 & 2.00 &  \\ 
  11/E3 &  & B & 5.00 & 1.45 & 3.00 & 4.50 & 1.50 & 2.50 &  \\ 
  11/E4 &  & B & 11.00 & 3.09 & 4.00 & 5.00 & 1.32 & 2.00 &  \\ 
  11/E4 & M-PSI/07 & B & 11.00 & 3.09 & 2.50 & 3.00 & 1.32 & 2.00 &  \\ 
  12/A1 &  & NB & 1.00 & 12.00 & 2.00 & 2.00 & 8.00 & 1.00 &  \\ 
  12/B1 &  & NB & 1.00 & 15.00 & 6.00 & 1.00 & 8.00 & 3.00 &  \\ 
  12/B2 &  & NB & 1.00 & 21.00 & 9.00 & 1.00 & 12.50 & 5.00 &  \\ 
  12/C1 &  & NB & 2.00 & 21.00 & 4.00 & 1.05 & 15.00 & 3.00 &  \\ 
  12/C2 &  & NB & 2.00 & 19.00 & 6.00 & 2.00 & 8.00 & 5.00 &  \\ 
  12/D1 &  & NB & 1.00 & 17.00 & 3.00 & 1.00 & 9.00 & 1.00 &  \\ 
  12/D1 & IUS/09 & NB & 1.00 & 8.00 & 0.00 &  &  &  & O \\ 
  12/D2 &  & NB & 1.00 & 22.00 & 7.00 & 1.00 & 15.00 & 6.00 &  \\ 
  12/E1 &  & NB & 1.00 & 19.00 & 4.00 & 1.00 & 12.00 & 3.00 &  \\ 
  12/E2 &  & NB & 1.00 & 17.00 & 3.00 & 2.00 & 11.50 & 2.00 &  \\ 
  12/E3 &  & NB & 1.00 & 15.00 & 5.00 & 2.00 & 14.00 & 4.00 &  \\ 
  12/F1 &  & NB & 1.25 & 23.75 & 13.00 & 1.00 & 17.00 & 7.00 &  \\ 
  12/G1 &  & NB & 2.00 & 17.00 & 6.00 & 1.00 & 11.50 & 2.00 &  \\ 
  12/G2 &  & NB & 1.00 & 20.00 & 7.00 & 2.00 & 11.00 & 4.00 &  \\ 
  12/H1 &  & NB & 2.00 & 13.00 & 2.50 & 1.00 & 6.00 & 1.00 &  \\ 
  12/H2 &  & NB & 2.00 & 16.00 & 2.00 & 2.00 & 7.00 & 1.00 &  \\ 
  12/H3 &  & NB & 3.00 & 17.00 & 2.00 & 2.00 & 12.00 & 1.00 &  \\ 
  13/A1 &  & NB & 0.00 & 12.11 & 1.50 & 0.00 & 11.00 & 2.00 & Oo \\ 
  13/A2 &  & NB & 1.00 & 13.00 & 0.50 & 0.00 & 13.16 & 1.00 & o \\ 
  13/A3 &  & NB & 0.00 & 13.00 & 0.00 & 0.00 & 14.50 & 1.00 & OOo* \\ 
  13/A4 &  & NB & 1.00 & 16.00 & 0.00 & 1.00 & 18.00 & 0.00 & Oo* \\ 
  13/A5 &  & NB & 0.00 & 16.00 & 6.00 & 0.00 & 14.00 & 5.00 & Oo \\ 
  13/B1 &  & NB & 3.00 & 9.00 & 0.00 & 3.00 & 10.00 & 0.00 & Oo* \\ 
  13/B2 &  & NB & 2.00 & 15.00 & 0.00 & 2.00 & 14.00 & 0.00 & Oo \\ 
  13/B3 &  & NB & 1.00 & 14.00 & 0.00 & 1.50 & 11.55 & 0.00 & Oo \\ 
  13/B4 &  & NB & 1.00 & 9.00 & 0.00 & 1.00 & 8.00 & 0.00 & Oo \\ 
  13/B4 & SECS-P/09 & NB &  &  &  & 1.00 & 5.00 & 0.00 & o \\ 
  13/B5 &  & NB & 0.00 & 21.00 & 0.00 & 1.00 & 17.00 & 0.00 & OOo \\ 
  13/C1 &  & NB & 1.00 & 15.00 & 0.00 & 1.00 & 14.50 & 0.00 & Oo \\ 
  13/C1 & SECS-P/04 & NB & 0.50 & 15.00 & 0.00 & 0.00 & 14.50 & 0.00 & Ooo \\ 
  13/D1 &  & NB & 0.00 & 17.00 & 1.00 & 0.00 & 15.00 & 1.00 & Oo \\ 
  13/D1 & SECS-S/02 & NB &  &  &  & 0.00 & 11.00 & 1.00 & o \\ 
  13/D2 &  & NB & 0.00 & 11.00 & 0.00 & 0.00 & 15.00 & 0.00 & OOoo* \\ 
  13/D3 &  & NB & 0.00 & 16.00 & 0.00 & 0.00 & 21.50 & 0.00 & OOoo* \\ 
  13/D4 &  & NB & 0.00 & 12.00 & 2.00 & 0.00 & 10.00 & 1.00 & Oo \\ 
  14/A1 &  & NB & 2.00 & 17.00 & 1.00 & 2.00 & 13.50 & 0.00 & o \\ 
  14/A2 &  & NB & 2.00 & 15.50 & 2.00 & 1.05 & 14.00 & 1.00 &  \\ 
  14/B1 &  & NB & 2.00 & 15.00 & 0.00 & 2.00 & 10.00 & 0.00 & Oo \\ 
  14/B1 & SPS/03 & NB &  &  &  & 0.50 & 10.00 & 0.00 & o \\ 
  14/B2 &  & NB & 2.00 & 11.00 & 1.00 & 1.00 & 13.00 & 1.00 &  \\ 
  14/B2 & SPS/05 & NB & 2.00 & 11.00 & 0.00 & 1.00 & 13.00 & 0.00 & Oo \\ 
  14/B2 & SPS/06 & NB &  &  &  & 1.00 & 13.00 & 0.00 & o \\ 
  14/C1 &  & NB & 2.00 & 17.00 & 1.00 & 2.00 & 11.00 & 1.00 &  \\ 
  14/C2 &  & NB & 3.00 & 17.00 & 1.00 & 2.85 & 15.00 & 1.00 &  \\ 
  14/D1 &  & NB & 2.00 & 16.00 & 2.00 & 2.00 & 12.00 & 2.00 &  \\ \bottomrule
\end{longtable}
\end{scriptsize}

\end{document}